\documentclass[draftcls,onecolumn]{IEEEtran}
\usepackage{graphicx,amsmath,amssymb,array,stfloats,cuted,midfloat,psfrag}
\usepackage[noadjust]{cite}
\newtheorem{theorem}{\textbf{Theorem}}[section]
\newtheorem{definition}{\textbf{Definition}}[section]
\newtheorem{lemma}{\textbf{Lemma}}[section]
\newtheorem{corollary}{\textbf{Corollary}}[section]
\newtheorem{claim}{\textbf{Claim}}[section]
\begin{document}
\title{Successive Wyner-Ziv Coding Scheme and its Application to the Quadratic Gaussian CEO Problem}
\author{Jun Chen,~\IEEEmembership{Member,~IEEE}, Toby Berger, ~\IEEEmembership{Fellow,~IEEE}
\thanks{Jun Chen and Toby Berger are supported in part by NSF Grant
CCR-033 0059 and a grant from the National Academies Keck Futures
Initiative (NAKFI).}}
\date{}
\markboth{} {Chen \MakeLowercase{\textit{et al.}}} \maketitle

\begin{abstract}
We introduce a distributed source coding scheme called successive
Wyner-Ziv coding. We show that any point in the rate region of the
quadratic Gaussian CEO problem can be achieved via the successive
Wyner-Ziv coding. The concept of successive refinement in the
single source coding is generalized to the distributed source
coding scenario, which we refer to as distributed successive
refinement. For the quadratic Gaussian CEO problem, we establish a
necessary and sufficient condition for distributed successive
refinement, where the successive Wyner-Ziv coding scheme plays an
important role.
\end{abstract}


\IEEEpeerreviewmaketitle

\section{Introduction}

The problem of distributed source coding has assumed renewed
interest in recent years. Many practical compression schemes have
been proposed for Slepian-Wolf coding (e.g.
\cite{Schongberg,Coleman} and the reference therein) and Wyner-Ziv
coding (e.g. \cite{Cheng} and the reference therein), whose
performances are close to the fundamental theoretical bounds
\cite{Slepian,Wyner}. Therefore it is of interest to reduce the
general distributed source coding problem to these well-studied
cases.

Given $L$ i.i.d. discrete sources $X_1, X_2, \cdots,X_L$, the
Slepian-Wolf rate region is the union of all the rate vectors
$(R_1,R_2,\cdots,R_L)$ satisfying
\begin{eqnarray}
\sum\limits_{i\in\mathcal{A}}R_i\geq
H(X_{\mathcal{A}}|X_{\mathcal{I}_L\backslash\mathcal{A}}),\quad
\forall \mbox{ nonempty set } \mathcal{A}\subseteq\mathcal{I}_L,
\end{eqnarray}
where $\mathcal{I}_L=\{1,2,\cdots,L\}$ and
$X_{\mathcal{A}}=\{X_i\}_{i\in\mathcal{A}}$. The Slepian-Wolf
reigon is a contra-polymatroid \cite{Edmonds,Tse} with $L!$
vertices. Specifically, if $\pi$ is a permutation on
$\mathcal{I}_L$, define the vector
$(R_1(\pi),R_2(\pi),\cdots,R_L(\pi))$ by
\begin{eqnarray}
R_{\pi (i)}(\pi) &=& H(X_{\pi (i)}|X_{\pi (i+1)},\cdots,X_{\pi
(L)}),\quad i = 1,\cdots,L-1,\\
R_{\pi (L)}(\pi) &=& H(X_{\pi (L)}).
\end{eqnarray}
Then $(R_1(\pi),R_2(\pi),\cdots,R_L(\pi))$ is a vertex of the
Slepian-Wolf region for every permutation $\pi$. It is known that
vertices of the Slepian-Wolf region can be achieved with a
complexity which is significantly lower than that of a general
point. It was observed in \cite{Rimoldi} that by splitting a
source into two virtual sources one can reduce the problem of
coding an arbitrary point in a $L$-dimensional Slepian-Wolf region
to that of coding a vertex of a ($2L-1$)-dimensional Slepian-Wolf
region. The source-splitting approach was also adopted in
distributed lossy source coding \cite{Zamir}.  In the distributed
lossy source coding scenario, we shall refer to source splitting
as quantization splitting (from the encoder viewpoint) or
description refinement (from the decoder viewpoint) since it is
the quantization output, not the source, that gets split. Finally
we want to point out that the source-splitting idea has a dual in
the problem of coding for multiple access channels, which is
referred to as rate-splitting
\cite{Carleial,Urbanke,Grant,Rimoldi2}.


The rest of this paper is divided into 3 sections. In Section II,
we introduce a low complexity successive Wyner-Ziv coding schemem
and prove that any point in the rate region of the quadratic
Gaussian CEO problem can be achieved via this scheme. The duality
between the superposition coding in multiaccess communication and
the successive Wyner-Ziv coding is briefly discussed. The concept
of distributed successive refinement is introduced in Section III.
The quadratic Gaussian CEO problem is used as an example, for
which the necessary and sufficient condition for the distributed
successive refinement is established. We conclude the paper in
Section IV.

In this paper, we use boldfaced letters to indicate
($n$-dimensional) vectors, capital letters for random objects, and
small letters for their realizations. For example, we let
$\mathbf{X}=(X(1),\cdots,X(n))^T$ and
$\mathbf{x}=(x(1),\cdots,x(n))^T$. Calligraphic letters are used
to indicate a set (say, $\mathcal{A}$). We use $U_{\mathcal{A}}$
to denote the vector $(U_i)_{i\in\mathcal{A}}$ with index $i$ in
an increasing order and use $U_{\mathcal{A},\mathcal{B}}$ to
denote $(U_{\mathcal{A},j})_{j\in\mathcal{B}}$ \footnote {Here the
elements of $\mathcal{A}$ and $\mathcal{B}$ are assumed to be
nonnegative integers.}. For example, if
$\mathcal{A}=\mathcal{B}=\{1,2\}$, then
$U_{\mathcal{A}}=(U_1,U_2)$ and
$U_{\mathcal{A},\mathcal{B}}=(U_{1,1},U_{2,1},U_{1,2},U_{2,2})$.
Here $U_i$ (and $U_{i,j}$) can be a random variable, a constant or
a function. We let $U_{\mathcal{A}}$ be a constant if
$\mathcal{A}$ is an empty set. We use $\mathcal{I}_K$ to denote
the set $\{1,2,\cdots,K\}$ for any positive integer $K$.

\section{Successive Wyner-Ziv Coding Scheme}

In this paper, we adopt the model of the CEO problem. But some of
our results also hold for many other distributed source coding
models. The CEO problem has been studied for many years
\cite{Pinsker,Gray,BergerCEO}. Here is a brief description of this
problem (also see Fig. 1).

\begin{figure}[hbt] \label{fig1}
\centering
\begin{psfrags}
\small
\psfrag{X}[c]{$X(t)$}%
\psfrag{y1}[l]{$Y_1(t)$} %
\psfrag{y2}[l]{$Y_2(t)$} %
\psfrag{yl}[l]{$Y_L(t)$} %
\psfrag{dots}[c]{$\vdots$} %
\psfrag{Obser-}[l]{Obser-} %
\psfrag{vations}[l]{vations} %
\psfrag{en1}[c]{Encoder $1$} %
\psfrag{en2}[c]{Encoder $2$} %
\psfrag{enl}[c]{Encoder $L$} %
\psfrag{r1}[l]{$R_1$} %
\psfrag{r2}[l]{$R_2$} %
\psfrag{rl}[l]{$R_L$} %
\psfrag{decoder}[c]{Decoder} %
\psfrag{Xhat}[c]{$\hat X(t)$} %
\includegraphics[scale=0.8]{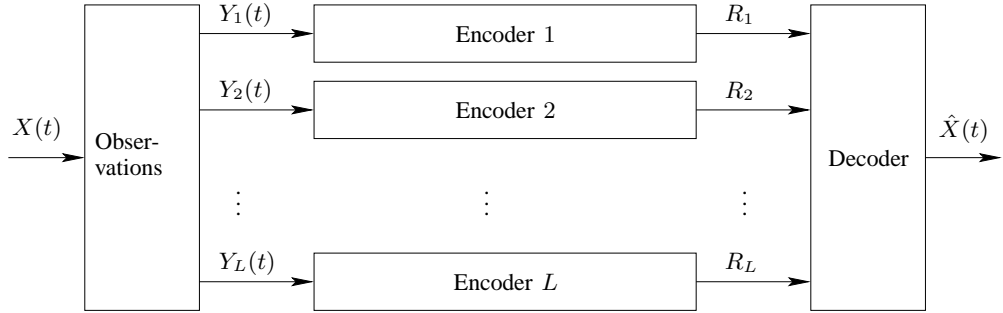}
\end{psfrags}
\caption{Model of the CEO problem}
\end{figure}

Let $\{X(t), Y_1(t), \cdots, Y_L(t)\}^{\infty}_{t=1}$ be a
temporally memoryless source with instantaneous joint probability
distribution $P(x,y_1,\cdots,y_L)$ on
$\mathcal{X}\times\mathcal{Y}_1\times\cdots\times\mathcal{Y}_L$,
where $\mathcal{X}$ is the common alphabet of the random variables
$X(t)\  \mbox{for } t=1, 2, \cdots$, and $\mathcal{Y}_i\ \ (i=1,
2, \cdots, L)$ is the common alphabet of the random variables
$Y_i(t)\ \mbox{for } t=1, 2, \cdots$. $\{X(t)\}_{t=1}^{\infty}$ is
the target data sequence that the decoder is interested in. This
data sequence cannot be observed directly. $L$ encoders are
deployed, where encoder $i$ observes
$\{Y_i(t)\}_{t=1}^{\infty},\mbox{ }i=1,2,\cdots,L$. The data rate
at which encoder $i\mbox{ } (i = 1, 2,\cdots,L)$ may communicate
information about its observations to the decoder is limited to
$R_i$ bits per second. The encoders are not permitted to
communicate with each other. Finally, the decision $\{\hat
X(t)\}_{t=1}^{\infty}$ is computed from the combined data at the
decoder so that a desired fidelity can be satisfied.

\begin{definition}
An $L$-tuple of rates $R_{\mathcal{I}_L}$ is said to be
$D$-admissible if for all $\epsilon
> 0$, there exists an $n_0$ such that for all $n>n_0$ there exist
encoders:
\begin{eqnarray*}
 f_{i}^{(n)}: \mathcal{Y}_i^n \rightarrow \left\{1,2,\cdots,\lfloor
 2^{n(R_i+\epsilon)}\rfloor\right\},\quad i=1,2,\cdots,L,
\end{eqnarray*}
and a decoder:
\begin{equation*}
g^{(n)}: \left\{1,2,\cdots,\lfloor
 2^{n(R_1+\epsilon)}\rfloor\right\}\times \left\{1,2,\cdots,\lfloor
 2^{n(R_2+\epsilon)}\rfloor\right\}
\cdots \times \left\{1,2,\cdots,\lfloor
 2^{n(R_L+\epsilon)}\rfloor\right\} \rightarrow \mathcal{X}^n,
\end{equation*}
such that
\begin{equation*}
\frac{1}{n}\mathbb{E}\left[\sum\limits_{t=1}^{n} d\left(X(t), \hat
X(t)\right)\right] \leq D+\epsilon,
\end{equation*}
where $\mathbf{\hat X} = g^{(n)}\left(f^{(n)}_1
(\mathbf{Y}_1),\cdots,f^{(n)}_L(\mathbf{Y}_M)\right)$ and
$d(\cdot,\cdot) : \mathcal{X}\times\mathcal{X}\rightarrow [0,
d_{\max}]$ is a given distortion measure. We use $\mathcal{R}(D)$
to denote the set of all $D$-admissible rate tuples.
\end{definition}

\begin{definition}[Berger-Tung rate region]
Let
\begin{equation}
\mathcal{R}(W_{\mathcal{I}_L})=\left\{R_{\mathcal{I}_L}:\sum\limits_{i\in\mathcal{A}}R_i\geq
I\left(Y_{\mathcal{A}};W_{\mathcal{A}}|W_{\mathcal{A}^c}\right),\forall
\mbox{ nonempty set }\mathcal{A}\subseteq\mathcal{I}_L \right\}
\end{equation}
where $W_i\rightarrow Y_i\rightarrow (X,Y_{\mathcal{I}_L\backslash
\{i\}},W_{\mathcal{I}_L\backslash \{i\}})$ form a Markov chain for
all $i\in\mathcal{I}_L$. The Berger-Tung rate region with respect
to distortion $D$ is
\begin{equation}
\mathcal{R}_{BT}(D)=\mbox{conv}\left(\bigcup\limits_{W_{\mathcal{I}_L}\in\mathcal{W}(D)}\mathcal{R}(W_{\mathcal{I}_L})\right),
\end{equation}
where $\mathcal{W}(D)$ is the set of all $W_{\mathcal{I}_L}$
satisfying the following properties:
\begin{itemize}
\item [(i)] $W_i\rightarrow Y_i\rightarrow (X,Y_{\mathcal{I}_L\backslash
\{i\}},W_{\mathcal{I}_L\backslash \{i\}})$ form a Markov chain for
all $i\in\mathcal{I}_L$.
\item [(ii)] There exists a function
\begin{equation*}
f:\mathcal{W}_1\times\cdots\times\mathcal{W}_L\rightarrow\mathcal{X}
\end{equation*}
such that $Ed(X,\hat X)\leq D$, where $\hat
X=f(W_{\mathcal{I}_L})$.
\end{itemize}
\end{definition}

It was shown in \cite{Berger,Tung,Housewright} that
$\mathcal{R}_{BT}\subseteq\mathcal{R}(D)$. The Berger-Tung rate
region is the largest known achievable rate region for the general
CEO problem although it was shown by K\"{o}rner and Marton
\cite{Korner} that it is not always tight. Computing the
Berger-Tung rate region involves complicated optimization and
convexification. Hence we shall only focus on
$\mathcal{R}(W_{\mathcal{I}_L})$. We will see that for the
quadratic Gaussian CEO problem, the properties of the Berger-Tung
rate region are determined completely by those of
$\mathcal{R}(W_{\mathcal{I}_L})$.

It was proved in \cite{Viswanath,ChenCEO} that
$\mathcal{R}(W_{\mathcal{I}_L})$ is a contra-polymatroid with $L!$
vertices. Specifically, if $\pi$ is a permutation on
$\mathcal{I}_L$, define the vector $R_{\mathcal{I}_L}(\pi)$ by
\begin{eqnarray}
R_{\pi (i)}(\pi)&=&I(Y_{\pi (i)};W_{\pi (i)}|W_{\pi
(i+1)},\cdots,W_{\pi
(L)}),\quad i=1,\cdots,L-1,\\
R_{\pi (L)}(\pi)&=&I(Y_{\pi (L)};W_{\pi (L)}).
\end{eqnarray}
Then $R_{\mathcal{I}_L}(\pi)$ is a vertex of
$\mathcal{R}(W_{\mathcal{I}_L})$ for every permutation $\pi$. The
dominant face of $\mathcal{R}(W_{\mathcal{I}_L})$ is the convex
polytope consisting of all points
$R_{\mathcal{I}_L}\in\mathcal{R}(W_{\mathcal{I}_L})$ such that
$\sum_{i=1}^LR_i=I(Y_{\mathcal{I}_L};W_{\mathcal{I}_L})$. Any rate
tuple $R_{\mathcal{I}_L}$ on the dominant face of
$\mathcal{R}(W_{\mathcal{I}_L})$ has the property that
\begin{eqnarray*}
R'_{\mathcal{I}_L}\leq R_{\mathcal{I}_L}\Rightarrow
R'_{\mathcal{I}_L}=R_{\mathcal{I}_L},\quad\forall
R'_{\mathcal{I}_L}\in\mathcal{R}(W_{\mathcal{I}_L})
\end{eqnarray*}
where $R'_{\mathcal{I}_L}\leq R_{\mathcal{I}_L}$ means $R'_i\leq
R_i\mbox{ for all } i\in\mathcal{I}_L$. It is easy to check that
the vertices of $\mathcal{R}(W_{\mathcal{I}_L})$ are on its
dominant face. For each vertex $R_{\mathcal{I}_L}(\pi)$, there
exists a low-complexity successive Wyner-Ziv coding scheme which
can be roughly described as follows:
\begin{itemize}
\item [(i)] Encoder $\pi (L)$ employs conventional lossy source coding.
Encoder $\pi (i)\mbox{ } (i = L-1,L-2,\cdots,1)$ employs Wyner-Ziv
coding with side information $\mathbf{W}_{\pi
(i+1)},\cdots,\mathbf{W}_{\pi (L)}$ at decoder.
\item [(ii)] Decoder first
decodes the codeword $\mathbf{W}_{\pi (L)}$ from encoder $\pi
(L)$, then successively decodes the codeword $\mathbf{W}_{\pi
(i)}\mbox{ }(i =L-1,L-2,\cdots,1)$ from encoder $\pi (i)$ with
side information $\mathbf{W}_{\pi (i+1)},\cdots,\mathbf{W}_{\pi
(L)}$.
\end{itemize}
Rate tuples on the dominant face other than these $L!$ vertices
were previously known to be attainable only by one of two methods.
The first method known to achieve these difficult rate tuples was
time sharing between vertices. This approach can require as many
as $L$ successive decoding schemes\footnote{By
Carath$\acute{\mbox{e}}$odory's fundamental theorem
\cite{Eggleston}, any point in the convex closure of a connected
commpact set $\mathcal{A}$ in a $d$-dimensional Euclidean space
can be represented as a convex combination of $d+1$ or fewer
points in the original set $\mathcal{A}$.}, each scheme requiring
$L$ decoding steps. The second approach to achieve these rate
tuples is joint decoding of all users. This is very difficult to
implement in practice since random codes have a decoding
complexity of the order of
$2^{nI(Y_{\mathcal{I}_L};W_{\mathcal{I}_L})}$, where $n$ is the
block length.

We will show that any rate tuple in
$\mathcal{R}(W_{\mathcal{I}_L})$ can be achieved by a
low-complexity successive Wyner-Ziv coding scheme with at most
$2L-1$ steps. Without loss of generality, we only need to consider
the rate tuple on the dominant face of
$\mathcal{R}(W_{\mathcal{I}_L})$. Before proceeding to prove this
result, we shall first give a formal description of the general
successive Wyner-Ziv coding scheme.

Let
$(W_{1,\mathcal{I}_{m_1}},W_{2,\mathcal{I}_{m_2}},\cdots,W_{L,\mathcal{I}_{m_L}})$
jointly distributed with the generic source variables
$(X,Y_{\mathcal{I}_L})$ such that
$W_{i,\mathcal{I}_{m_i}}\rightarrow Y_i\rightarrow
(X,Y_{\mathcal{I}_L\backslash \{i\}},W_{j,\mathcal{I}_{m_j}},
j\in\mathcal{I}_L\backslash \{i\})$ form a Markov chain for all
$i\in\mathcal{I}_L$. Let $\sigma$ be a permutation on
$\{W_{1,\mathcal{I}_{1}},\cdots,W_{L,\mathcal{I}_{m_L}}\}$ such
that for all $i\in\mathcal{I}_L$, $W_{i,j}$ is placed before
$W_{i,k}$ if $j<k$ (we refer to this type of permutation as the
well-ordered permutation). Let $\{W_{i,j}\}_{\sigma}^-$ denote all
the random variables that appear before $W_{i,j}$ in the
permutation $\sigma$.

\textit{Random Binning at Encoder $i$}: In what follows we shall
adopt the notation and conventions of \cite{Cover}. Let
$n$-vectors $\mathbf{W}_{i,1}(1),\cdots,\mathbf{W}_{i,1}(M_{i,1})$
be drawn independently according to a uniform distribution over
the set $T_\epsilon(W_{i,1})$ of $\epsilon$-typical $W_{i,1}$
$n$-vectors, where $M_{i,1}=\lfloor 2^{n(
I(Y_i,W_{i,1})+\epsilon'_{i,1})}\rfloor$. That is,
$P(\mathbf{W}_{i,1}(k)=\mathbf{w}_{i,1})=1/\parallel
T_\epsilon(W_{i,1})\parallel$, if $\mathbf{w}_{i,1}\in
T_\epsilon(W_{i,1})$, and $=0$ otherwise. Distribute these vectors
into $N_{i,1}$ bins: $B_{i,1}(1),\cdots,B_{i,1}(N_{i,1})$, such
that
\begin{equation*}
\left\lfloor\frac{M_{i,1}}{N_{i,1}}\right\rfloor\leq|B_{i,1}(b)|_{W_{i,1}}\leq\left\lceil\frac{M_{i,1}}{N_{i,1}}\right\rceil,\quad
b=1,2,\cdots,N_{i,1},
\end{equation*}
where $N_{i,1}=\lfloor
2^{(nI(Y_i,W_{i,1}|\{W_{i,1}\}_{\sigma}^-)+\epsilon_{i,1})}\rfloor$
and $|B_{i,1}(b)|_{W_{i,1}}$ denotes the number of
$\mathbf{W}_{i,1}$-vectors in $B_{i,1}(b)$.

Successively from $j=2, j=3,\cdots$, to $j=m_i$, for each vector
$(k_{1},\cdots,k_{j-1})$ with $k_s\in\{1,2,\cdots,M_{i,s}\}\mbox{
}(s=1,\cdots,j-1)$, let
$\mathbf{W}_{i,j}(k_{1},\cdots,k_{j-1},1),\cdots,\mathbf{W}_{i,j}(k_{1},\cdots,k_{j-1},M_{i,j})$
be drawn i.i.d. according to a uniform distribution over the set
$T_\epsilon(W_{i,j}|\mathbf{w}_{i,1}(k_{1}),\cdots,\mathbf{w}_{i,j-1}(k_1,\cdots,k_{j-1}))$
of conditionally $\epsilon$-typical $\mathbf{w}_{i,j}$'s,
conditioned on $\mathbf{w}_{i,1}(k_{1})$,
$\cdots,\mathbf{w}_{i,j-1}(k_1,\cdots,k_{j-1})$, and distribute
them uniformly into $N_{i,j}$ bins:
$B_{i,j}(1),\cdots,B_{i,j}(N_{i,j})$ such that
\begin{equation*}
\left\lfloor\frac{M_{i,j}}{N_{i,j}}\right\rfloor\leq|B_{i,j}(b)|_{W_{i,j}}\leq\left\lceil\frac{M_{i,j}}{N_{i,j}}\right\rceil,\quad
b=1,2,\cdots,N_{i,j}.
\end{equation*}
Here $M_{i,j}=\lfloor 2^{n(
I(Y_i,W_{i,j}|W_{i,\mathcal{I}_{j-1}})+\epsilon'_{i,j})}\rfloor,
N_{i,j}=\lfloor
2^{n(I(Y_i,W_{i,j}|\{W_{i,j}\}_{\sigma}^-)+\epsilon_{i,j})}\rfloor$.
Note: $\epsilon_{i,j},\epsilon'_{i,j} (i\in\mathcal{I}_L,
j\in\mathcal{I}_{m_i})$ are positive numbers of the same order as
$\epsilon$ which can be made arbitrarily small as
$n\rightarrow\infty$. Furthermore, we require
$\epsilon_{i,j}>\epsilon'_{i,j}$ for all $i\in\mathcal{I}_L,
j\in\mathcal{I}_{m_i}$

\textit{Encoding at Encoder $i$}: Given a
$\mathbf{y}_i\in\mathcal{Y}^n_i$, find, if possible, a vector
$(k^*_{i,1},\cdots,k^*_{i,m_i})$ such that
\begin{equation*}
\left(\mathbf{y}_i,\mathbf{w}_{i,1}(k^*_{i,1}),\mathbf{w}_{i,2}(k^*_{i,1},k^*_{i,2}),\cdots,\mathbf{w}_{i,m_i}(k^*_{i,1},\cdots,k^*_{i,m_i})\right)\in
T_\epsilon(Y_i,W_{i,1},W_{i,2},\cdots,W_{i,m_i}).
\end{equation*}
Then find bins
$B_{i,1}(b^*_{i,1}),B_{i,2}(b^*_{i,2}),\cdots,B_{i,m_i}(b^*_{i,m_i})$
such that $B_{i,j}(b^*_{i,j})$ contains
$\mathbf{w}_{i,j}(k^*_{i,1},\cdots,k^*_{i,j}), j=1,2,\cdots,m_i$.
Send $(b^*_{i,1},\cdots,b^*_{i,m_i})$ to the decoder. If no such
$(k^*_{i,1},\cdots,k^*_{i,m_i})$ exists, simply send
$(0,\cdots,0)$.

We can see the resulting transmission rate of encoder $i$ is
\begin{equation}
R_i=\frac{1}{n}\log\left(\prod\limits_{j=1}^{m_i}N_{i,j}+1\right)\leq\sum\limits_{j=1}^{m_i}I(Y_i,W_{i,j}|\{W_{i,j}\}_{\sigma}^-)+\sum\limits_{j=1}^{m_i}\epsilon_{i,j}+\frac{1}{n}.\label{888}
\end{equation}

\textit{Decoding}: Given $(b^*_{i,1},\cdots,b^*_{i,m_i})$ for all
$i\in\mathcal{I}_L$, if
$(b^*_{i,1},\cdots,b^*_{i,m_i})=(0,\cdots,0)$ for some $i$,
declare a decoding failure. Otherwise decode as follows:

Let $\sigma(j)$ denote the $j^{th}$ element in permutation
$\sigma$. Let $s_1(j), s_2(j)$ be the first and second subscript
of $\sigma(j)$, respectively. For example, if $\sigma(j)=W_{3,2}$,
then $s_1(j)=3, s_2(j)=2$. Decoder first finds
$\mathbf{w}_{s_1(1),s_2(1)}(\hat k_{s_1(1),s_2(1)})$ in
$B(b^*_{s_1(1),s_2(2)})$. Note: $s_2(1)=1$. Since
$B(b^*_{s_1(1),s_2(2)})$ contains at most one vector, we have
$\hat k_{s_1(1),s_2(1)}=k^*_{s_1(1),s_2(1)}$. Successively from
$j=2, j=3,\cdots$, to $j=\sum_{i=1}^{L}m_i$, if in
$B_{s_1(j),s_2(j)}(b^*_{s_1(j),s_2(j)})$, there exists a unique
$\hat k_{s_1(j),s_2(j)}$ such that
\begin{eqnarray*}
\left(\mathbf{w}_{s_1(i),s_2(i)}(\hat k_{s_1(i),1},\hat
k_{s_1(i),2},\cdots,\hat
k_{s_1(i),s_2(i)}),i\in\mathcal{I}_{j}\right)\in
T_{\epsilon'}\left(W_{s_1(i),s_2(i)},i\in\mathcal{I}_j\right),
\end{eqnarray*}
decode $\mathbf{w}_{s_1(j),s_2(j)}(\hat k_{s_1(j),1},\hat
k_{s_1(j),2},\cdots,\hat k_{s_1(j),s_2(j)})$, otherwise declare a
decoding failure. Note: $\epsilon'$ is of the same order as
$\epsilon$ which can be made arbitrarily small as
$n\rightarrow\infty$.

By the standard technique, it can be shown that $Pr(\hat
k_{i,j}=k^*_{i,j}, \forall i\in\mathcal{I}_L,
j\in\mathcal{I}_{m_i})\rightarrow 1$ as $n\rightarrow\infty$.
Furthermore, by Markov Lemma \cite{Berger}, we have
\begin{equation*}
Pr\left(\left(\mathbf{X},
\mathbf{W}_{i,j}(k^*_{i,1},k^*_{i,2},\cdots,k^*_{i,j}),i\in\mathcal{I}_L,
j\in\mathcal{I}_{m_i}\right)\in T_{\epsilon'}\left(X,
W_{i,j},i\in\mathcal{I}_L,
j\in\mathcal{I}_{m_i}\right)\right)\rightarrow 1
\end{equation*}
as $n\rightarrow\infty$. Hence for any function
$g:\prod_{i=1}^L\prod_{j=1}^{m_i}\mathcal{W}_{i,j}\rightarrow\mathcal{X}$,
we have
\begin{equation*}
\frac{1}{n}\mathbb{E}\left[\sum\limits_{t=1}^n
d\left(X(t),g\left(W_{i,j}(k^*_{i,1},k^*_{i,2},\cdots,k^*_{i,j},t),i\in\mathcal{I}_L,
j\in\mathcal{I}_{m_i}\right)\right)\right]\leq
\mathbb{E}d\left(X,g(W_{i,j},i\in\mathcal{I}_L,
j\in\mathcal{I}_{m_i})\right)+\epsilon''d_{\max}
\end{equation*}
with high probability, where
$W_{i,j}(k^*_{i,1},k^*_{i,2},\cdots,k^*_{i,j},t)$ is the $t^{th}$
entry of $\mathbf{W}_{i,j}(k^*_{i,1},k^*_{i,2},\cdots,k^*_{i,j})$
and $\epsilon''$ is of the same order as $\epsilon$ which can be
made arbitrarily small as $n\rightarrow\infty$.

It is easy to see that if we let
$W'_{i,j}=W_{i,\mathcal{I}_j}\mbox{ }(\forall i\in\mathcal{I}_L,
j\in\mathcal{I}_{m_i})$, and replace $W_{i,j}$ by $W'_{i,j}$ in
(\ref{888}), $R_i$ is unaffected. Hence there is no loss of
generality to assume $W_{i,1}\rightarrow
W_{i,2}\rightarrow\cdots\rightarrow W_{i,m_i}\rightarrow Y_i,
\forall i\in\mathcal{I}_L$. We can view $W_{i,j}$ as a description
of $Y_i$, as $j$ gets larger, the description gets finer.

The above coding scheme can be interpreted in the following
intuitive way:

Encoder $i$ first splits $R_i$ into $m_i$ pieces:
$r_{i,j}=I(Y_i;W_{i,j}|\{W_{i,j}\}_{\sigma}^-), \forall
j\in\mathcal{I}_{m_i}$. Then successively from $j=1, j=2,\cdots$,
to $j=m_i$, it uses a Wyner-Ziv code with rate $r_{i,j}$ to convey
$\mathbf{W}_{i,j}$ to decoder which has the side information
$\{\mathbf{W}'_i\}_{\sigma}^-$. Decoder recovers
$\{\mathbf{W}_{i,j},j\in\mathcal{I}_{m_i},i\in\mathcal{I}_L\}$
successively according to the order in the permutation $\sigma$.
We can see that this scheme requires $\sum_{i=1}^L m_i$ Wyner-Ziv
coding steps. Thus we call it $\sum_{i=1}^L m_i$-successive
Wyner-Ziv coding scheme. A similar successive coding strategy was
developed in \cite{Draper} for tree-structured sensor networks.

The successive Wyner-Ziv encoding and decoding structure of the
above scheme significantly reduces the coding complexity compared
with joint decoding or time sharing scheme and makes the available
practical Wyner-Ziv coding techniques directly applicable to the
more general distributed source coding scenarios. Furthermore, the
successive Wyner-Ziv coding scheme has certain robust property
which is especially attractive in some applications. Since in the
successive Wyner-Ziv coding scheme, encoder $i$ essentially
transmits its codeword in $m_i$ packets. Each packet contains a
sub-codeword $\mathbf{W}_{i,j}\mbox{ } (j\in\mathcal{I}_{m_i})$.
If a packet, say packet $\mathbf{W}_{i,k}$ is lost in
transmission, the decoder is still able to decode packets
$\{\mathbf{W}_{i,k}\}_{\sigma}^-$. On the contrary, the jointly
decoding scheme does not possess this robust property since any
corruption in the transmitted codewords may cause a complete
failure in decoding.

We need introduce another definition before giving a formal
statement of our first theorem.
\begin{definition}\label{def23}
For any disjoint sets $\mathcal{A},
\mathcal{B}\subseteq\mathcal{I}_L$ ($\mathcal{A}$ is nonempty),
let
\begin{eqnarray*}
\mathcal{R}(W_{\mathcal{A}}|W_{\mathcal{B}},Z_{\mathcal{I}_L})=\left\{R_{\mathcal{A}}:\sum\limits_{i\in\mathcal{S}}R_i\geq
I\left(Y_{\mathcal{S}};W_{\mathcal{S}}|W_{\mathcal{A}\backslash\mathcal{S}},W_{\mathcal{B}},Z_{\mathcal{I}_L}\right),\forall
\mbox{ nonempty set }\mathcal{S}\subseteq\mathcal{A} \right\},
\end{eqnarray*}
where $Z_i\rightarrow W_{i}\rightarrow Y_i\rightarrow (X,
Y_{\mathcal{I}_L\backslash \{i\}},W_{\mathcal{I}_L\backslash
\{i\}},Z_{\mathcal{I}_L\backslash \{i\}})$ form a Markov chain for
all $i\in\mathcal{I}_L$.
\end{definition}

It's easy to check that
$\mathcal{R}(W_{\mathcal{A}}|W_{\mathcal{B}},Z_{\mathcal{I}_L})$
is a contra-polymatroid with $|\mathcal{A}|!$ vertices.
Specifically, if $\pi$ is a permutation on $\mathcal{A}$, define
the vector $R_{\mathcal{A}}(\pi)$ by
\begin{eqnarray}
R_{\pi (i)}(\pi)&=&I(Y_{\pi (i)};W_{\pi (i)}|W_{\pi
(i+1)},\cdots,W_{\pi (|\mathcal{A}|)},W_{\mathcal{B}},Z_{\mathcal{I}_L}),\quad i=1,\cdots,|\mathcal{A}|-1,\\
R_{\pi (|\mathcal{A}|)}(\pi)&=&I(Y_{\pi (|\mathcal{A}|)};W_{\pi
(|\mathcal{A}|)}|W_{\mathcal{B}},Z_{\mathcal{I}_L}).
\end{eqnarray}
Then $R_{\mathcal{A}}(\pi)$ is a vertex of
$\mathcal{R}(W_{\mathcal{A}}|W_{\mathcal{B}},Z_{\mathcal{I}_L})$
for every permutation $\pi$. The dominant face
$\mathcal{D}(W_{\mathcal{A}}|W_{\mathcal{B}},Z_{\mathcal{I}_L})$
of
$\mathcal{R}(W_{\mathcal{A}}|W_{\mathcal{B}},Z_{\mathcal{I}_L})$
is the convex polytope consisting of all points
$R_{\mathcal{A}}\in\mathcal{R}(W_{\mathcal{A}}|W_{\mathcal{B}},Z_{\mathcal{I}_L})$
such that
$\sum_{i\in\mathcal{A}}R_i=I(Y_{\mathcal{A}};W_{\mathcal{A}}|W_{\mathcal{B}},Z_{\mathcal{I}_L})$.
We have
$dim\left[\mathcal{D}(W_{\mathcal{A}}|W_{\mathcal{B}},Z_{\mathcal{I}_L})\right]\leq
|\mathcal{A}|-1$, where
$dim\left[\mathcal{D}(W_{\mathcal{A}}|W_{\mathcal{B}},Z_{\mathcal{I}_L})\right]$
is the dimension of $\mathcal{D}(W_{\mathcal{A}}|Z)$. The equality
holds only when the $|\mathcal{A}|!$ vertices are all distinct.
Any rate tuple
$R_{\mathcal{A}}\in\mathcal{D}(W_{\mathcal{A}}|W_{\mathcal{B}},Z_{\mathcal{I}_L})$
has the property that
\begin{equation*}
R'_{\mathcal{A}}\leq R_{\mathcal{A}}\Rightarrow
R'_{\mathcal{A}}=R_{\mathcal{A}},\quad\forall
R'_{\mathcal{A}}\in\mathcal{R}(W_{\mathcal{A}}|W_{\mathcal{B}},Z_{\mathcal{I}_L}).
\end{equation*}

\begin{theorem}\label{theorem21}
For any rate tuple
$R_{\mathcal{A}}\in\mathcal{D}(W_{\mathcal{A}}|W_{\mathcal{B}},Z_{\mathcal{I}_L})$,
there exist random variables
$(W'_{i,1},\cdots,W'_{i,m_i})_{i\in\mathcal{A}}$ jointly
distributed with $(X, Y_{\mathcal{I}_L},W_{\mathcal{I}_L},
Z_{\mathcal{I}_L})$ satisfying
\begin{itemize}
\item [(i)]
$(W'_{i,m_i})_{i\in\mathcal{A}}=W_{\mathcal{A}}$ (i.e.,
$(W'_{i,m_i})_{i\in\mathcal{A}}$ and $W_{\mathcal{A}}$ are just
two different names of the same random vector),
\item [(ii)] $\sum_{i\in\mathcal{A}} m_{i}\leq |\mathcal{A}|+dim\left[\mathcal{D}(W_{\mathcal{A}}|W_{\mathcal{B}},Z_{\mathcal{I}_L})\right]$ and $m_i\leq 2$ for all
$i\in\mathcal{A}$,
\item [(iii)]
$Z_{i}\rightarrow W'_{i,1}\rightarrow W'_{i,m_i}\rightarrow
Y_i\rightarrow (X, Y_{\mathcal{I}_L\backslash
\{i\}},W_{\mathcal{I}_L\backslash
\{i\}},Z_{\mathcal{I}_L\backslash
\{i\}},W'_{j,\mathcal{I}_{m_j}},j\in\mathcal{A}\backslash \{i\})$
form a Markov chain for all $i\in\mathcal{I}_L$,
\end{itemize}
and a well-ordered permutation $\sigma$ on
$\{W'_{i,\mathcal{I}_{m_i}},i\in\mathcal{A}\}$ such that
\begin{equation}
R_i=\sum\limits_{j=1}^{m_i}I\left(Y_i;W'_{i,j}|\{W'_{i,j}\}_{\sigma}^-,W_{\mathcal{B}},Z_{\mathcal{I}_L}\right),\quad
\forall\mbox{ }i\in\mathcal{A}. \label{rateconstraint}
\end{equation}
\end{theorem}
\begin{proof}
The theorem can be proved in a similar manner as in \cite{Grant}.
The details are omitted.
\end{proof}

When $\mathcal{A}=\mathcal{I}_L$ and $\mathcal{B}=\emptyset$,
Theorem \ref{theorem21} says that if $\mathbf{Z}_{\mathcal{I}_L}$
is available at the decoder, then encoders $1,2,\cdots,L$ can
convey $\mathbf{W}_{\mathcal{I}_L}$ to the decoder via a
$(2L-1)$-successive Wyner-Ziv coding scheme as long as
$R_{\mathcal{I}_L}\in\mathcal{R}(W_{\mathcal{I}_L}|Z_{\mathcal{I}_L})$.

It is noteworthy that $2L-1$ is just an upper bound, for the rate
tuple on the boundary of
$\mathcal{D}(W_{\mathcal{I}_L}|Z_{\mathcal{I}_L})$, the coding
complexity can be further reduced. For example, consider the case
where $L=3$. Let $V_1$ be the vertex corresponding to permutation
$\pi_1=(1,2,3)$, i.e.,
\begin{equation*}
V_1=(I(Y_1;W_{1}|Z_{\mathcal{I}_3},W_{2},W_{3}),I(Y_2;W_{2}|Z_{\mathcal{I}_3},W_{3}),I(Y_3;W_{3}|Z_{\mathcal{I}_3})).
\end{equation*}
Let $V_2$ be the vertex corresponding to permutation
$\pi_2=(1,3,2)$, i.e.,
\begin{equation*}
V_2=(I(Y_1;W_{1}|Z_{\mathcal{I}_3},W_{2},W_{3}),I(Y_2;W_{2}|Z_{\mathcal{I}_3}),I(Y_3;W_{3}|Z_{\mathcal{I}_3},W_{2})).
\end{equation*}
For any rate tuple $R_{\mathcal{I}_3}$ on the edge connecting
$V_1$ and $V_2$, we have
$R_1=I(Y_1;W_{1}|Z_{\mathcal{I}_3},W_{2},W_{3})$. Hence encoder 1
can use a Wyner-Ziv code to convey $\mathbf{W}_{1}$ to the decoder
if $(\mathbf{Z}_{\mathcal{I}_3},\mathbf{W}_2,\mathbf{W}_3)$ are
already available at the decoder. Since $(R_2,R_3)$ is on the
dominant face of $\mathcal{R}(W_{2},W_{3}|Z_{\mathcal{I}_3})$, by
Theorem \ref{theorem21}, decoder 2 and encoder 3 can convey
$(\mathbf{W}_{2},\mathbf{W}_{3})$ to the decoder via a
$3$-successive Wyner-Ziv coding scheme if
$\mathbf{Z}_{\mathcal{I}_3}$ is available to the decoder. Thus
overall it is a 4-successive Wyner-Ziv coding scheme.

In general we can imitate the approach in \cite{UrbankeM}. For
$\emptyset\subset\mathcal{A}\subset\mathcal{I}_L$, define the
hyperplane
\begin{equation*}
\mathcal{H}(\mathcal{A})=\left\{\mathcal{R}_{\mathcal{I}_L}\in\mathcal{R}^L:\sum\limits_{i\in\mathcal{A}}R_i=I(Y_{\mathcal{A}};W_{\mathcal{A}}|Z_{\mathcal{I}_L})\right\}
\end{equation*}
and let
$\mathcal{F}_{\mathcal{A}}=\mathcal{H}(\mathcal{A})\cap\mathcal{D}(W_{\mathcal{I}_L}|Z_{\mathcal{I}_L})$.
If
$\emptyset\subset\mathcal{A}_1\subset\mathcal{A}_2\subset\cdots\subset\mathcal{A}_k\subset\mathcal{I}_L$,
is a telescopic sequence of subsets, then
$\mathcal{F}_{\mathcal{A}_1}\cap\mathcal{F}_{\mathcal{A}_2}\cap\cdots\cap\mathcal{F}_{\mathcal{A}_k}$
is a face of $\mathcal{D}(W_{\mathcal{I}_L}|Z_{\mathcal{I}_L})$.
Conversely, every face of
$\mathcal{D}(W_{\mathcal{I}_L}|Z_{\mathcal{I}_L})$ can be written
in this form. Let
$\mathcal{B}_i=\mathcal{A}_i\backslash\mathcal{A}_{i-1}$,
$i=1,2,\cdots,k+1$, where we set $\mathcal{A}_0=\emptyset$ and
$\mathcal{A}_{k+1}=\mathcal{I}_L$. Let $\Xi$ be the set of
permutation $\pi$ on $\mathcal{I}_L$ such that
\begin{equation*}
\left\{\pi\left(\sum_{j=0}^{k-i} |\mathcal{B}_{k+1-j}|
+1\right),\cdots,\pi\left(\sum_{j=0}^{k+1-i}
|\mathcal{B}_{k+1-j}|\right)\right\}=\mathcal{B}_{i},\quad
i=1,2,\cdots,k+1.
\end{equation*}
Each permutation $\pi\in\Xi$ is associated with a vertex of
$\mathcal{F}_{\mathcal{A}_1}\cap\mathcal{F}_{\mathcal{A}_2}\cap\cdots\cap\mathcal{F}_{\mathcal{A}_k}$
and vice versa. Hence
$\mathcal{F}_{\mathcal{A}_1}\cap\mathcal{F}_{\mathcal{A}_2}\cap\cdots\cap\mathcal{F}_{\mathcal{A}_k}$
has totally $|\Xi|=\prod_{i=1}^{k+1}(|\mathcal{B}_i|!)$ vertices.
Moreover, we have
$dim(\mathcal{F}_{\mathcal{A}_1}\cap\mathcal{F}_{\mathcal{A}_2}\cap\cdots\cap\mathcal{F}_{\mathcal{A}_k})\leq
L-k-1$, where the equality holds if these $|\Xi|$ vertices are all
distinct. For any rate tuple
$R_{\mathcal{I}_L}\in\mathcal{F}_{\mathcal{A}_1}\cap\mathcal{F}_{\mathcal{A}_2}\cap\cdots\cap\mathcal{F}_{\mathcal{A}_k}$,
it is easy to verify that $\mathcal{R}_{\mathcal{B}_i}$ is on the
dominant face of
$\mathcal{R}(W_{\mathcal{B}_i}|W_{\bigcup_{j=1}^{i-1}\mathcal{B}_j},Z_{\mathcal{I}_L})$,
$i=1,2,\cdots,k+1$. Hence by successively applying Theorem
\ref{theorem21}, we can conclude that an
$(L+\overline{L})$-successive Wyner-Ziv coding scheme is
sufficient for conveying $\mathbf{W}_{\mathcal{I}_L}$ to the
decoder if it has the side information
$\mathbf{Z}_{\mathcal{I}_L}$, where
\begin{eqnarray*}
\overline{L}=\sum\limits_{i=1}^{k+1}dim\left[\mathcal{D}\left(W_{\mathcal{B}_i}|W_{\bigcup_{j=1}^{i-1}\mathcal{B}_j},Z_{\mathcal{I}_L}\right)\right]=dim(\mathcal{F}_{\mathcal{A}_1}\cap\mathcal{F}_{\mathcal{A}_2}\cap\cdots\cap\mathcal{F}_{\mathcal{A}_k}).
\end{eqnarray*}

\begin{corollary}\label{cor21}
Any rate tuple $R_{\mathcal{I}_L}$ on the dominant face of
$\mathcal{R}(W_{\mathcal{I}_L})$ can be achieved via a
$K$-successive Wyner-Ziv coding scheme for some $K\leq 2L-1$.
\end{corollary}
\begin{proof}
Apply Theorem 1 with $Z_{\mathcal{I}_L}$ being a constant.
\end{proof}

This successive Wyner-Ziv coding scheme has a dual in the multiple
access communication, which we call the successive superposition
coding scheme.

Consider an $L$-user discrete memoryless multiple-access channel.
This is defined in terms of a stochastic matrix
\begin{equation*}
W:\mathcal{X}\times\cdots\times\mathcal{X}_L\rightarrow
\mathcal{Y}
\end{equation*}
with entries $W(y|x_1,\cdots,x_L)$ describing the probability that
the channel output is $y$ when the inputs are $x_1,\cdots,x_L$.

Now we give a brief description of the successive superposition
coding scheme.

Let
$X_{1,\mathcal{I}_{m_1}},X_{2,\mathcal{I}_{m_2}},\cdots,X_{L,\mathcal{I}_{m_L}}$
be independent, i.e.,
\begin{eqnarray*}
p(x_{1,\mathcal{I}_{m_1}},x_{2,\mathcal{I}_{m_2}},\cdots,x_{L,\mathcal{I}_{m_L}})=p(x_{1,\mathcal{I}_{m_1}})p(x_{2,\mathcal{I}_{m_2}})\cdots
p(x_{L,\mathcal{I}_{m_L}}),
\end{eqnarray*}
and $x_{i,m_i}\in\mathcal{X}_i$ for all $i\in\mathcal{I}_L$. Let
$\sigma$ be a well-ordered permutation on the set
$\{X_{1,\mathcal{I}_{m_1}},X_{2,\mathcal{I}_{m_2}},\cdots,X_{L,\mathcal{I}_{m_L}}\}$.

Encoder $i$: Let $n$-vectors
$\mathbf{X}_{i,1}(1),\cdots,\mathbf{X}_{i,1}(M_{i,1})$ be drawn
independently according to the marginal distribution $p(x_{i,1})$,
where $M_{i,1}=\lceil
2^{n(I(X_{i,1};Y|\{X_{i,1}\}_{\sigma}^-)-\epsilon_{i,1})}\rceil$.
Successively from $j=2, j=3,\cdots$, to $j=m_i$, for each vector
$(k_{1},\cdots,k_{j-1})$ with $k_s\in\{1,2,\cdots,M_{i,s}\}\mbox{
}(s=1,\cdots,j-1)$, let
$\mathbf{X}_{i,j}(k_{1},\cdots,k_{j-1},1),\cdots,\mathbf{X}_{i,j}(k_{1},\cdots,k_{j-1},M_{i,j})$
be drawn i.i.d. according to the marginal conditional distribution
$p(x_{i,j}|x_{i,1},\cdots,x_{i,j-1})$, conditioned on
$\mathbf{x}_{i,1}(k_{1}),\cdots$,
$\mathbf{x}_{i,j-1}(k_{1},\cdots,k_{j-1})$. Here $M_{i,j}=\lceil
2^{n(I(X_{i,j};Y|\{X_{i,j}\}_{\sigma}^-)-\epsilon_{i,j})}\rceil$.
Only $x_{i,m_i}(k_1,\cdots,k_{m_i})$'s will be transmitted. Hence
the resulting rate for encoder $i$ is
\begin{equation}
R_i=\frac{1}{n}\log\left(\prod\limits_{j=1}^{m_i}M_{i,j}\right)\geq\sum\limits_{j=1}^{m_i}I(X_{i,j};Y|\{X_{i,j}\}_{\sigma}^-))-\sum\limits_{j=1}^{m_i}\epsilon_{i,j}.\label{999}
\end{equation}

Decoder: Suppose $x_{i,m_i}(k^*_1,\cdots,k^*_{m_i})$ is
transmitted, which generates $\mathbf{y}\in\mathcal{Y}^n$ at the
channel output. Decoder first finds a $\hat k_{s_1(1),s_2(1)}$
such that $\mathbf{y}$ and $\mathbf{x}_{s_1(1),s_2(1)}(\hat
k_{s_1(1),s_2(1)})$ are jointly typical. If there is no or more
than one such $\hat k_{s_1(1),s_2(1)}$, declare a decoding
failure. Otherwise proceed as follows:

Successively from $j=2, j=3, \cdots$ to $j=\sum_{i=1}^L m_i$, if
there exists a unique $\hat k_{s_1(j),s_2(j)}$ such that
\begin{eqnarray*}
\left(\mathbf{y}, \mathbf{x}_{s_1(i),s_2(i)}(\hat
k_{s_1(i),1},\hat k_{s_1(i),2},\cdots,\hat
k_{s_1(i),s_2(i)}),i\in\mathcal{I}_{j}\right)\in
T_\epsilon\left(Y, X_{s_1(i),s_2(i)},i\in\mathcal{I}_j\right),
\end{eqnarray*}
decode $\mathbf{x}_{s_1(j),s_2(j)}(\hat k_{s_1(j),1},\hat
k_{s_1(j),2},\cdots,\hat k_{s_1(j),s_2(j)})$, otherwise declare a
decoding failure.

By the standard technique, it can be shown that $Pr(\hat
k_{i,j}=k^*_{i,j}, \forall i\in\mathcal{I}_L,
j\in\mathcal{I}_{m_i})\rightarrow 1$ as $n\rightarrow\infty$.

It is easy to see that if we let
$X'_{i,j}=X_{i,\mathcal{I}_j}\mbox{ }(\forall i\in\mathcal{I}_L,
j\in\mathcal{I}_{m_i})$, and replace $X_{i,j}$ by $X'_{i,j}$ in
(\ref{999}), $R_i$ is unaffected. Hence there is no loss of
generality to assume $X_{i,1}\rightarrow
X_{i,2}\rightarrow\cdots\rightarrow X_{i,m_i}\rightarrow (Y,
X_{j,\mathcal{I}_{m_j}},j\in\mathcal{I}_L\backslash\{i\})$ for all
$i\in\mathcal{I}_L$. With this Markov structure, this scheme can
be understood more intuitively since we can think that along this
Markov chain, high rate codebook is successively generated via
superposition on low rate codebook. We refer to the above coding
scheme as $\sum_{i=1}^L m_i$-successively superposition coding.

Our successive superposition coding scheme is similar to the
rate-splitting scheme introduced in \cite{Grant}. Actually every
rate-splitting scheme can be converted into a successive
superposition scheme. To see this, for each user $i$, let $f_i$ be
a splitting function such that
$X_i=f_i(U_{i,1},U_{i,2},\cdots,U_{i,m_i})$ and let
$X_{i,m_i}=X_i, X_{i,j}=U_{i,\mathcal{I}_j}, j=1,2,\cdots,m_i-1$,
$i\in\mathcal{I}_L$. Then $X_{i,1}\rightarrow
X_{i,2}\rightarrow\cdots\rightarrow X_{i,m_i}\rightarrow (Y,
X_{j,\mathcal{I}_{m_j}},j\in\mathcal{I}_L\backslash\{i\})$ form a
Markov chain for all $i\in\mathcal{I}_L$. In \cite{Grant}
$U_{i,1},U_{i,2},\cdots,U_{i,m_i}$ are required to be
independent\footnote{This independence condition is unnecessary
since $U_{i,1},U_{i,2},\cdots,U_{i,m_i}$ are all controlled by
user $i$. But this condition facilitates the codebook construction
and storage since now the high-rate codebook at each user is
essentially a product of low-rate codebooks.}, if we remove this
condition, then every successive superposition coding scheme can
also be converted into a rate-splitting scheme by simply setting
$U_{i,j}=X_{i,j}, \forall j\in\mathcal{I}_{m_i}$ and
$f_i(U_{i,1},U_{i,2},\cdots,U_{i,m_i})=U_{i,m_i}$.

Let
\begin{equation*}
\mathcal{R}(X_{\mathcal{I}_L})=\left\{R_{\mathcal{I}_L}\in\mathbb{R}^L_+:\sum\limits_{i\in\mathcal{A}}R_i\leq
I\left(X_{\mathcal{A}};Y|X_{\mathcal{I}_L\backslash\mathcal{A}}\right),\forall
\mbox{ nonempty set }\mathcal{A}\subseteq\mathcal{I}_L \right\}.
\end{equation*}
Ahlswede \cite{Ahlswede} and Liao \cite{Liao} proved that
\begin{equation*}
\mathcal{C}=\mbox{conv}\left(\bigcup\limits_{p(x_1)p(x_2)\cdots
p(x_L)}\mathcal{R}(X_{\mathcal{I}_L})\right),
\end{equation*}
where $\mathcal{C}$ is the capacity region of the synchronous
channel.

It can be shown that if $p(x_1,x_2\cdots,x_L)=p(x_1)p(x_2)\cdots
p(x_L)$, then $\mathcal{R}(X_{\mathcal{I}_L})$ is a polymatroid
\cite{Edmonds}\cite{Tse} with $L!$ vertices. Specifically, if
$\pi$ is a permutation on $\mathcal{I}_L$, define the vector
$R_{\mathcal{I}_L}(\pi)$ by
\begin{eqnarray}
R_{\pi (i)}(\pi)&=&I(X_{\pi (i)}(\pi);Y|X_{\pi
(1)}(\pi),\cdots,X_{\pi
(i-1)}(\pi)),\quad i=1,\cdots,L-1,\\
R_{\pi (L)}(\pi)&=&I(X_{\pi (L)}(\pi);Y).
\end{eqnarray}
Then $R_{\mathcal{I}_L}(\pi)$ is a vertex of
$\mathcal{R}(X_{\mathcal{I}_L})$ for every permutation $\pi$. The
dominant face of $\mathcal{R}(X_{\mathcal{I}_L})$ is the convex
polytope consisting of all points
$R_{\mathcal{I}_L}\in\mathcal{R}(X_{\mathcal{I}_L})$ such that
$\sum_{i=1}^LR_i=I(X_{\mathcal{I}_L};Y)$. Any rate tuple
$R_{\mathcal{I}_L}$ on the dominant face of
$\mathcal{R}(X_{\mathcal{I}_L})$ has the property that
\begin{equation*}
R'_{\mathcal{I}_L}\geq R_{\mathcal{I}_L}\Rightarrow
R'_{\mathcal{I}_L}=R_{\mathcal{I}_L},\quad\forall
R'_{\mathcal{I}_L}\in\mathcal{R}(X_{\mathcal{I}_L}).
\end{equation*}

The following corollary is a dual result of Corollary \ref{cor21}.
The proof is similar to that of Corollary \ref{cor21} and thus
omitted.
\begin{corollary}
Any rate tuple $R_{\mathcal{I}_L}$ on the dominant face of
$\mathcal{R}(\mathcal{I}_L)$ can be achieved via a $K$-successive
superposition coding scheme for some $K\leq 2L-1$.
\end{corollary}

Although we assumed discrete-alphabet sources and bounded
distortion measure in the previous discussion, all our results can
be extended to the Gaussian case with squared distortion measure
along the lines of \cite{Wynerinfo,OohamaG,Oohama}. Now we proceed
to study the quadratic Gaussian CEO problem \cite{Viswanathan},
for which some stronger conclusions can be drawn. Let
$\{X(t)\}_{t=1}^\infty$ are i.i.d. Gaussian random variables with
zero mean and variance $\sigma^2_X$. Let
$\{Y_i(t)\}_{t=1}^{\infty} =\{X(t) + N_i(t)\}_{t=1}^\infty$ for
all $i\in\mathcal{I}_L$, where $\{N_i(t)\}_{t=1}^\infty$ are i.i.d
Gaussian random variables independent of $\{X(t)\}_{t=1}^\infty$
with mean zero and variance $\sigma^2_{N_i}$ . Also, the random
processes $\{N_k(t)\}_{t=1}^\infty$ and $\{N_j(t)\}_{t=1}^\infty$
are independent for $j\neq k$. For each $i\in\mathcal{I}_L$, let
$W_i=Y_i+T_i$. where $T_i\sim\mathcal{N}(0,\sigma^2_{T_i})$ is
independent of $(X,Y_{I_L},T_{\mathcal{I}_L\backslash\{i\}})$.
Moreover, let
\begin{equation}
r_i=I(Y_i,W_i|X)=\frac{1}{2}\log\frac{\sigma^2_{N_i}+\sigma^2_{T_i}}{\sigma^2_{T_i}},\quad
\forall\mbox{ }i\in\mathcal{I}_L.\label{14}
\end{equation}
It was computed in \cite{ChenCEO,TseCEO} that
\begin{eqnarray}
\mathcal{R}(W_{\mathcal{I}_L})&=&\left\{R_{\mathcal{I}_L}:\sum\limits_{i\in\mathcal{A}}R_i\geq\frac{1}{2}\log\left(\frac{\frac{1}{\sigma^2_X}+\sum\limits_{i=1}^L\frac{1-\exp(-2r_i)}{\sigma^2_{N_i}}}{\frac{1}{\sigma^2_X}+\sum\limits_{i\in\mathcal{A}^c}\frac{1-\exp(-2r_i)}{\sigma^2_{N_i}}}\right)
+\sum\limits_{i\in\mathcal{A}}r_i, \forall\mbox{ nonempty set
}\mathcal{A}\subseteq\mathcal{I}_L\right\}\\&\triangleq&\mathcal{R}(r_{\mathcal{I}_L}).\label{17}
\end{eqnarray}
Furthermore, it was shown in \cite{OohamaU,TseCEO} that
\begin{equation}
\mathcal{R}(D)=\bigcup\limits_{r_{\mathcal{I}_L}\in\mathcal{F}(D)}\mathcal{R}(r_{\mathcal{I}_L}),\label{90}
\end{equation}
where
\begin{equation}
\mathcal{F}(D)=\left\{r_{\mathcal{I}_L}\in\mathbb{R}_+^L:\frac{1}{\sigma^2_X}+\sum\limits_{i=1}^L\frac{1-\exp(-2r_i)}{\sigma^2_{N_i}}\geq\frac{1}{D}\right\}.\label{8}
\end{equation}

\begin{definition}\label{def24}
Let $\partial\mathcal{R}(D)$ denote the boundary of
$\mathcal{R}(D)$, i.e.,
\begin{eqnarray*}
\partial\mathcal{R}(D)=\{R_{\mathcal{I}_L}\in\mathcal{R}(D): R'_{\mathcal{I}_L}\leq
R_{\mathcal{I}_L}\Rightarrow R'_{\mathcal{I}_L}=R_{\mathcal{I}_L},
\mbox{ for all }R'_{\mathcal{I}_L}\in\mathcal{R}(D) \}.
\end{eqnarray*}
\end{definition}

Clearly, any rate tuple inside $\mathcal{R}(D)$ is dominated by
some rate tuple in $\partial\mathcal{R}(D)$. Therefore there is no
loss of generality to focus on $\partial\mathcal{R}(D)$.

Now we proceed to compute $\partial\mathcal{R}(D)$ for the
quadratic Gaussian CEO problem. The closed-form expression of
$\partial\mathcal{R}(D)$ is hard to get. Instead, we shall
characterize the supporting hyperplanes of $\mathcal{R}(D)$, since
the upper envelope of their union is exactly
$\partial\mathcal{R}(D)$. The supporting hyperplanes of
$\mathcal{R}(D)$ have the following parametric form:
\begin{eqnarray}
\sum\limits_{i=1}^L\alpha_iR_i=\varphi(\alpha_{\mathcal{I}_L}),
\end{eqnarray}
where $\alpha_{\mathcal{I}_L}$ is a unit ($l_2$-norm) vector in
$\mathbb{R}^L_+$ and
\begin{eqnarray}
\varphi(\alpha_{\mathcal{I}_L})=\min\limits_{R_{\mathcal{I}_L}\in\mathcal{R}(D)}\sum\limits_{i=1}^L\alpha_iR_i.
\end{eqnarray}
Since $\mathcal{R}(r_{\mathcal{I}_L})$ is a contra-polymatroid, by
\cite[Lemma 3.3]{Tse}, a solution to the optimization problem
\begin{eqnarray}
\min\sum\limits_{i=1}^L\alpha_iR_i\quad\mbox{subject to}\quad
R_{\mathcal{I}_L}\in\mathcal{R}(r_{\mathcal{I}_L})
\end{eqnarray}
is attained at a vertex $R_{\mathcal{I}_L}(\pi^*)$ where is
$\pi^*$ any permutation such that
$\alpha_{\pi^*(1)}\geq\cdots\geq\alpha_{\pi^*(L)}$. That is,
\begin{eqnarray*}
\min\limits_{R_{\mathcal{I}_L}\in\mathcal{R}(r_{\mathcal{I}_L})}\sum\limits_{i=1}^L\alpha_iR_i&=&\sum\limits_{i=1}^L\alpha_iR_i(\pi^*)
=\sum\limits_{i=1}^{L-1}\left(\left(\alpha_{\pi^*(i)}-\alpha_{\pi^*(i+1)}\right)\sum\limits_{j=1}^iR_{\pi^*(j)}(\pi^*)\right)+\alpha_{\pi^*(L)}\sum\limits_{i=1}^LR_{\pi^*(i)}(\pi^*)
\\&=&\sum\limits_{i=1}^{L-1}\left(\alpha_{\pi^*(i)}-\alpha_{\pi^*(i+1)}\right)\left(\frac{1}{2}\log\left(\frac{\frac{1}{\sigma^2_X}+\sum\limits_{j=1}^L\frac{1-\exp(-2r_j)}{\sigma^2_{N_j}}}{\frac{1}{\sigma^2_X}+\sum\limits_{j=i+1}^{L}\frac{1-\exp(-2r_{\pi^*(j)})}{\sigma^2_{N_{\pi^*(j)}}}}\right)
+\sum\limits_{j=1}^ir_{\pi^*(j)}\right)\\
&&+\alpha_{\pi^*(L)}\left(\frac{1}{2}\log\left(1+\sigma^2_X\sum\limits_{j=1}^L\frac{1-\exp(-2r_j)}{\sigma^2_{N_j}}\right)+\sum\limits_{j=1}^Lr_j\right).
\end{eqnarray*}
Hence we have
\begin{eqnarray}
\varphi(\alpha_{\mathcal{I}_L})=\min\limits_{r_{\mathcal{I}_L}\in\mathbb{R}^L_+}&&\sum\limits_{i=1}^{L-1}\left(\alpha_{\pi^*(i)}-\alpha_{\pi^*(i+1)}\right)\left(\frac{1}{2}\log\left(\frac{\frac{1}{\sigma^2_X}+\sum\limits_{j=1}^L\frac{1-\exp(-2r_j)}{\sigma^2_{N_j}}}{\frac{1}{\sigma^2_X}+\sum\limits_{j=i+1}^{L}\frac{1-\exp(-2r_{\pi^*(j)})}{\sigma^2_{N_{\pi^*(j)}}}}\right)
+\sum\limits_{j=1}^ir_{\pi^*(j)}\right)\nonumber\\
&&+\alpha_{\pi^*(L)}\left(\frac{1}{2}\log\left(1+\sigma^2_X\sum\limits_{j=1}^L\frac{1-\exp(-2r_j)}{\sigma^2_{N_j}}\right)+\sum\limits_{j=1}^Lr_j\right)\label{23}
\end{eqnarray}
subject to
\begin{equation}
\frac{1}{\sigma^2_X}+\sum\limits_{i=1}^L\frac{1-\exp(-2r_i)}{\sigma^2_{N_i}}\geq\frac{1}{D}.\label{24}
\end{equation}
Since we can decrease $r_{\pi^*(1)}$ to make the constraint in
(\ref{24}) tight and keep the sum in (\ref{23}) decreasing at the
same time (If $r_{\pi^*(1)}$ attains 0 but the constraint in
(\ref{24}) is still not tight, then apply the same procedure to
$r_{\pi^*(2)}$ and so on.), we can rewrite (\ref{23}) and
(\ref{24}) as
\begin{eqnarray}
\varphi(\alpha_{\mathcal{I}_L})=\min\limits_{r_{\mathcal{I}_L}\in\mathbb{R}^L_+}&&\sum\limits_{i=1}^{L-1}\left(\alpha_{\pi^*(i)}-\alpha_{\pi^*(i+1)}\right)\left(\sum\limits_{j=1}^ir_{\pi^*(j)}-\frac{1}{2}\log\left(\frac{D}{\sigma^2_X}+\sum\limits_{j=i+1}^{L}\frac{D-D\exp(-2r_{\pi^*(j)})}{\sigma^2_{N_{\pi^*(j)}}}\right)
\right)\nonumber\\
&&+\alpha_{\pi^*(L)}\left(\frac{1}{2}\log\frac{\sigma^2_X}{D}+\sum\limits_{j=1}^Lr_j\right)
\end{eqnarray}
subject to
\begin{eqnarray}
\frac{1}{\sigma^2_X}+\sum\limits_{i=1}^L\frac{1-\exp(-2r_i)}{\sigma^2_{N_i}}=\frac{1}{D}.\label{equalityconstraint}
\end{eqnarray}
Let $r^*_{\mathcal{I}_L}$ be the minimizer of the above
optimization problem. Introduce Lagrange multipliers
$\lambda_{\mathcal{I}_L}\in\mathbb{R}^L$ for the inequality
constraints $r_{\mathcal{I}_L}\in\mathbb{R}^L_+$ and a multiplier
$\nu\in\mathbb{R}$ for the equality constraint
(\ref{equalityconstraint}). Define
\begin{eqnarray}
G(r_{\mathcal{I}_L},\lambda_{\mathcal{I}_L},\nu)&=&\sum\limits_{i=1}^{L-1}\left(\alpha_{\pi^*(i)}-\alpha_{\pi^*(i+1)}\right)\left(\sum\limits_{j=1}^ir_{\pi^*(j)}-\frac{1}{2}\log\left(\frac{D}{\sigma^2_X}+\sum\limits_{j=i+1}^{L}\frac{D-D\exp(-2r_{\pi^*(j)})}{\sigma^2_{N_{\pi^*(j)}}}\right)
\right)\nonumber\\
&&+\alpha_{\pi^*(L)}\left(\frac{1}{2}\log\frac{\sigma^2_X}{D}+\sum\limits_{j=1}^Lr_j\right)-\sum\limits_{i=1}^L\lambda_ir_i-\nu\left(\frac{1}{\sigma^2_X}+\sum\limits_{i=1}^L\frac{1-\exp(-2r_i)}{\sigma^2_{N_i}}\right).
\end{eqnarray}
We obtain the \textit{Karush-Kuhn-Tucker} conditions \cite{Boyd}
\begin{eqnarray*}
r^*_k&\geq& 0,\quad \lambda_k\geq 0,\quad \lambda_kr^*_k=0,\quad
k=1,2,\cdots,L,\quad\quad\quad\frac{1}{\sigma^2_X}+\sum\limits_{k=1}^L\frac{1-\exp(-2r^*_k)}{\sigma^2_{N_k}}=\frac{1}{D},\\
\left.\frac{\partial G}{\partial r_{\pi^*(1)}}\right|_{r_{\pi^*(1)}=r^*_{\pi^*(1)}}&=&-\frac{2\nu\exp(-2r^*_{\pi^*(1)})}{\sigma^2_{N_{\pi^*(1)}}}+\alpha_{\pi^*(1)}-\lambda_{\pi^*(i)}=0,\\
\left.\frac{\partial G}{\partial
r_{\pi^*(k)}}\right|_{r_{\pi^*(k)}=r^*_{\pi^*(k)}}&=&-\frac{\exp(-2r^*_{\pi^*(k)})}{\sigma^2_{N_{\pi^*(k)}}}\sum\limits_{i=1}^{k-1}(\alpha_{\pi^*(i)}-\alpha_{\pi^*(i+1)})\left(\frac{1}{\sigma^2_X}+\sum\limits_{j=i+1}^L\frac{1-\exp(-2r^*_{\pi^*(j)})}{\sigma^2_{N_{\pi^*(j)}}}\right)^{-1}\nonumber\\
&&+\alpha_{\pi^*(k)}-\lambda_{\pi^*(k)}-\frac{2\nu\exp(-2r^*_{\pi^*(k)})}{\sigma^2_{N_{\pi^*(k)}}}\\
&=&-\frac{\exp(-2r^*_{\pi^*(k)})}{\sigma^2_{N_{\pi^*(k)}}}\sum\limits_{i=1}^{k-1}(\alpha_{\pi^*(i)}-\alpha_{\pi^*(i+1)})\left(\frac{1}{D}-\sum\limits_{j=1}^i\frac{1-\exp(-2r^*_{\pi^*(j)})}{\sigma^2_{N_{\pi^*(j)}}}\right)^{-1}
\\&&+\alpha_{\pi^*(k)}-\lambda_{\pi^*(k)}-\frac{2\nu\exp(-2r^*_{\pi^*(k)})}{\sigma^2_{N_{\pi^*(k)}}}\\
&=&0,\quad k=2,3,\cdots,L.
\end{eqnarray*}
By the complementary slackness condition, i.e., $\lambda_k>0
\Rightarrow r^*_k=0$, we can solve these equations to get
\begin{eqnarray}
r^*_{\pi^*(1)}&=&\left[\frac{1}{2}\log\frac{2\nu}{\alpha_{\pi^*(1)}\sigma^2_{N_{\pi^*(1)}}}\right]^+,\label{30}\\
r^*_{\pi^*(k)}&=&\left[\frac{1}{2}\log\left(\frac{2\nu+\sum\limits_{i=1}^{k-1}(\alpha_{\pi^*(i)}-\alpha_{\pi^*(i+1)})\left(\frac{1}{D}-\sum\limits_{j=1}^i\frac{1-\exp(-2r^*_{\pi^*(j)})}{\sigma^2_{N_{\pi^*(j)}}}\right)^{-1}}{\alpha_{\pi^*(k)}\sigma^2_{N_{\pi^*(k)}}}\right)\right]^+,
k=2,3,\cdots,L \label{31}
\end{eqnarray}
where $\nu$ is uniquely determined by the distortion constraint
\begin{equation}
\frac{1}{\sigma^2_X}+\sum\limits_{i=1}^L\frac{1-\exp(-2r^*_i)}{\sigma^2_{N_i}}=\frac{1}{D}\label{dc1}
\end{equation}
and $r^*_{\mathcal{I}_L}$ can be computed recursively from
$r^*_{\pi^*(1)}$, $r^*_{\pi^*(2)}, \cdots,$ to $r^*_{\pi^*(L)}$.

In the above we assume $\alpha_i>0$ for all $i\in\mathcal{I}_L$.
Now suppose
$\alpha_{\pi^*(i)}\geq\cdots\geq\alpha_{\pi^*(\widetilde{L})}>0=\alpha_{\pi^*(\widetilde{L}+1)}=\cdots=\alpha_{\pi^*(L)}$.
We can let
$r^*_{\pi^*(\widetilde{L}+1)}=\cdots=r^*_{\pi^*(L)}=\infty$. If
\begin{equation}
\frac{1}{\sigma^2_X}+\sum\limits_{i=\widetilde{L}+1}^L\frac{1}{\sigma^2_{N_{\pi^*(i)}}}>\frac{1}{D}\label{dc2},
\end{equation}
then we have $r^*_{\pi^*(1)}=\cdots=r^*_{\pi^*(\widetilde{L})}=0$
and correspondingly $\varphi(\alpha_{\mathcal{I}_L})=0$.
Otherwise, solve
$r^*_{\pi^*(1)},\cdots,r^*_{\pi^*(\widetilde{L})}$ from (\ref{30})
(\ref{31}) with the distortion constraint (\ref{dc1}) replaced by
\begin{equation}
\frac{1}{\sigma^2_X}+\sum\limits_{i=1}^{\widetilde{L}}\frac{1-\exp(-2r^*_{\pi^*(i)})}{\sigma^2_{N_{\pi^*(i)}}}+\sum\limits_{i=\widetilde{L}+1}^L\frac{1}{\sigma^2_{N_{\pi^*(i)}}}=\frac{1}{D}.\label{dc11}
\end{equation}

Let $\mathcal{T}(\alpha_{\mathcal{I}_L},D)$ with $\alpha_i>0$
$(\forall i\in\mathcal{I}_L)$ be a supporting hyperplane of
$\mathcal{R}(D)$. By (\ref{90}), we have
\begin{equation*}
\mathcal{T}(\alpha_{\mathcal{I}_L},D)\cap\partial\mathcal{R}(D)=\mathcal{T}(\alpha_{\mathcal{I}_L},D)\cap\mathcal{R}(D)=\mathcal{T}(\alpha_{\mathcal{I}_L},D)\cap\left(\bigcup\limits_{r_{\mathcal{I}_L}\in\mathcal{F}(D)}\mathcal{R}(r_{\mathcal{I}_L})\right).
\end{equation*}
If
$\mathcal{T}(\alpha_{\mathcal{I}_L},D)\cap\mathcal{R}(r_{\mathcal{I}_L})\neq\emptyset$
for some $r_{\mathcal{I}_L}\in\mathcal{F}(D)$, then we must have
$R_{\mathcal{I}_L}(\pi^*)\in\mathcal{T}(\alpha_{\mathcal{I}_L},D)\cap\mathcal{R}(r_{\mathcal{I}_L})$,
where $R_{\mathcal{I}_L}(\pi^*)$ is a vertex (associated with
permutation $\pi^*$) of $\mathcal{R}(r_{\mathcal{I}_L})$. Now it
follows by the above Lagrangian optimization that
$\mathcal{R}(r_{\mathcal{I}_L})=\mathcal{R}(r^*_{\mathcal{I}_L})$.
Therefore, we have
\begin{equation*}
\mathcal{T}(\alpha_{\mathcal{I}_L},D)\cap\partial\mathcal{R}(D)=\mathcal{T}(\alpha_{\mathcal{I}_L},D)\cap\mathcal{R}(r^*_{\mathcal{I}_L}).
\end{equation*}
Clearly,
$\mathcal{T}(\alpha_{\mathcal{I}_L},D)\cap\mathcal{R}(r^*_{\mathcal{I}_L})$
is a face of the dominant face of
$\mathcal{R}(r^*_{\mathcal{I}_L})$. Let
$(\mathcal{B}'_1,\cdots,\mathcal{B}'_{k})$ be a partition of
$\mathcal{I}_L$ such that $\alpha_m=\alpha_n$ for any
$\alpha_m,\alpha_n\in\mathcal{B}'_i$ $(i=1,2,\cdots,k)$ and
$\alpha_m>\alpha_n$ for any $\alpha_m\in\mathcal{B}'_i,
\alpha_n\in\mathcal{B}'_j$ $(i<j)$. Let $\Xi'$ be the set of
permutation $\pi$ on $\mathcal{I}_L$ such that
\begin{equation*}
\left\{\pi\left(\sum\limits_{j=1}^{i-1}|\mathcal{B}'_j|+1\right),\cdots,\pi\left(\sum\limits_{j=1}^{i}|\mathcal{B}'_j|\right)\right\}=\mathcal{B}'_i,\quad
i=1,2,\cdots,k.
\end{equation*}
$\mathcal{T}(\alpha_{\mathcal{I}_L},D)\cap\mathcal{R}(r^*_{\mathcal{I}_L})$
has totally $|\Xi'|=\prod_{i=1}^{k}(|\mathcal{B}'_i|!)$ vertices,
each of which is associated with a permutation $\pi\in\Xi'$.
Furthermore,
$dim(\mathcal{T}(\alpha_{\mathcal{I}_L},D)\cap\mathcal{R}(r^*_{\mathcal{I}_L}))\leq
L-k$, where the equality holds if these $|\Xi'|$ vertices are all
distinct. Finally, we want to point out that if
$\alpha_1=\cdots=\alpha_L$, then
$\mathcal{T}(\alpha_{\mathcal{I}_L},D)\cap\mathcal{R}(r^*_{\mathcal{I}_L})$
is the minimum sum-rate region of $\mathcal{R}(D)$ \cite{ChenCEO}.



\begin{corollary}
For the quadratic Gaussian CEO problem, any rate tuple
$R_{\mathcal{I}_L}\in\partial\mathcal{R}(D)$ can be achieved via a
$K$-successive Wyner-Ziv coding scheme for some $K\leq 2L-1$.
\end{corollary}
\begin{proof}
Since
$\mathcal{R}(D)=\bigcup_{r_{\mathcal{I}_L}\in\mathcal{F}(D)}\mathcal{R}(r_{\mathcal{I}_L})$,
for any rate tuple $R_{\mathcal{I}_L}\in\partial\mathcal{R}(D)$,
there exists a vector $r_{\mathcal{I}_L}\in\mathcal{F}(D)$ such
that $R_{\mathcal{I}_L}\in\mathcal{R}(r_{\mathcal{I}_L})$.
Furthermore, by Definition \ref{def24}, it's easy to see that
$R_{\mathcal{I}_L}$ must be on the dominant face of
$\mathcal{R}(r_{\mathcal{I}_L})$. Now the desired result follows
from Corollary \ref{cor21}.

To get detailed information about the coding complexity of a rate
tuple $R_{\mathcal{I}_L}\in\partial\mathcal{R}(D)$, we can proceed
as follows. Let $\mathcal{T}(\alpha_{\mathcal{I}_L},D)$ be the
supporting hyperplane of $\partial\mathcal{R}(D)$ such that
$R_{\mathcal{I}_L}\in\mathcal{T}(\alpha_{\mathcal{I}_L},D)\cap\partial\mathcal{R}(D)$.
Use the Lagrangian optimization method to find
$\mathcal{R}(r^*_{\mathcal{I}_L})$ with
$r^*_{\mathcal{I}_L}\in\mathcal{F}(D)$ such that
$\mathcal{T}(\alpha_{\mathcal{I}_L},D)\cap\partial\mathcal{R}(D)=\mathcal{T}(\alpha_{\mathcal{I}_L},D)\cap\mathcal{R}(r^*_{\mathcal{I}_L})$.
Let
$\mathcal{F}\subseteq\mathcal{T}(\alpha_{\mathcal{I}_L},D)\cap\mathcal{R}(r^*_{\mathcal{I}_L})$
be the lowest dimensional face of
$\mathcal{R}(r^*_{\mathcal{I}_L})$ that contains
$R_{\mathcal{I}_L}$. We can conclude that $R_{\mathcal{I}_L}$ is
achievable via an $(L+dim(\mathcal{F}))$-successive Wyner-Ziv
coding scheme.
\end{proof}


\section{Distributed Successive Refinement}
In the previous section, we have shown that the successive
Wyner-Ziv coding scheme suffices to achieve any rate tuple on the
boundary of rate region for the quadratic Gaussian CEO problem. We
shall extend this result to the multistage source coding scenario.
\begin{definition}
For $R_{\mathcal{I}_L,1}\leq R_{\mathcal{I}_L,2}\leq\cdots\leq
R_{\mathcal{I}_L,M}$ and $D_1\geq D_2\geq\cdots\geq D_M$, we say
the $M$-stage source coding
\begin{equation*}
(R_{\mathcal{I}_L,1},D_1)\nearrow
(R_{\mathcal{I}_L,2},D_2)\nearrow\cdots\nearrow
(R_{\mathcal{I}_L,M},D_M)
\end{equation*}
is feasible if for any $\epsilon >0$, there exists an $n_0$ such
that for $n>n_0$ there exist encoders:
\begin{equation*}
f_{i,j}^{(n)}:\mathcal{Y}_i^n\rightarrow\left\{1,2,\cdots,\lfloor
2^{n(R_{i,j}-R_{i,j-1}+\epsilon)}\rfloor\right\},\quad
i=1,2,\cdots,L,\mbox{ }j=1,2,\cdots,M,
\end{equation*}
and decoders:
\begin{eqnarray*}
g_j^{(n)}:\prod\limits_{k=1}^j\prod\limits_{i=1}^L\left\{1,2,\cdots,\lfloor
2^{n(R_{j,k}-R_{j,k-1}+\epsilon)}\rfloor\right\}\rightarrow\mathcal{X}^n,\quad
j=1,2,\cdots,M,
\end{eqnarray*}
such that
\begin{eqnarray*}
\frac{1}{n}\mathbb{E}\left[\sum\limits_{t=1}^nd\left(X(t),\hat
X_j(t)\right)\right]\leq D_j+\epsilon,\quad j=1,2,\cdots,M,
\end{eqnarray*}
where
\begin{equation*}
\mathbf{\hat
X}_j=g_j^{(n)}\left(f_{1,1}^{(n)}(\mathbf{Y}_1),\cdots,f_{L,1}^{(n)}(\mathbf{Y}_L),\cdots,f_{1,j}^{(n)}(\mathbf{Y}_1),\cdots,f_{L,j}^{(n)}(\mathbf{Y}_L)\right),\quad
j=1,2,\cdots,M.
\end{equation*}
Here we assume $R_{\mathcal{I}_L,0}=(0,\cdots,0)$.
\end{definition}

The following definition can be viewed as a natural generalization
of the successive refinement in the single source coding
\cite{Koshelev1,Koshelev2,Equitz,RimoldiS} to the distributed
source coding scenario.
\begin{definition}[Distributed Successive Refinement]
Let $D^*(R_{\mathcal{I}_L}) = \min\{D :
R_{\mathcal{I}_L}\in\mathcal{R}(D)\}$. For
$R_{\mathcal{I}_L,1}\leq R_{\mathcal{I}_L,2}\leq\cdots\leq
R_{\mathcal{I}_L,M}$, we say there exists an $M$-stage distributed
successive refinement scheme from $R_{\mathcal{I}_L,1}$ to
$R_{\mathcal{I}_L,2}$, to $\cdots\cdots$, to $R_{\mathcal{I}_L,M}$
if the $M$-stage source coding
\begin{equation*}
(R_{\mathcal{I}_L,1},D^*(R_{\mathcal{I}_L,1}))\nearrow
(R_{\mathcal{I}_L,2},D^*(R_{\mathcal{I}_L,2}))\nearrow
\cdots\nearrow (R_{\mathcal{I}_L,M},D^*(R_{\mathcal{I}_L,M}))
\end{equation*}
is feasible.
\end{definition}

\begin{theorem}\label{thm31}
For $R_{\mathcal{I}_L,1}\leq R_{\mathcal{I}_L,2}\leq\cdots\leq
R_{\mathcal{I}_L,M}$ and $D_1\geq D_2\geq\cdots\geq D_M$, the
$M$-stage source coding
\begin{equation*}
(R_{\mathcal{I}_L,1},D_1)\nearrow
(R_{\mathcal{I}_L,2},D_2)\nearrow\cdots\nearrow
(R_{\mathcal{I}_L,M},D_M)
\end{equation*}
is feasible if there exist random variables
$W_{\mathcal{I}_L,\mathcal{I}_M}$ jointly distributed with the
generic source variables $(X,Y_{\mathcal{I}_L})$ such that
\begin{equation*}
\left(R_{\mathcal{I}_L,j}-R_{\mathcal{I}_L,j-1}\right)\in\mathcal{R}(W_{\mathcal{I}_L,j}|W_{\mathcal{I}_L,j-1}),
\end{equation*}
where $W_{\mathcal{I}_L,\mathcal{I}_M}$ satisfy the following
properties:
\begin{itemize}
\item [(i)] $W_{i,1}\rightarrow W_{i,2}\rightarrow\cdots\rightarrow W_{i,M}\rightarrow Y_i\rightarrow
(X,Y_{\mathcal{I}_L\backslash\{i\}},W_{\mathcal{I}_L\backslash\{i\},\mathcal{I}_M})$
form a Markov chain for all $i\in\mathcal{I}_L$;
\item [(ii)] For each $j\in\mathcal{I}_M$, there exists a function $\hat X_j:\prod_{i=1}^L\mathcal{W}_{i,j}\rightarrow\mathcal{X}$ such that $\mathbb{E}d(X,\hat X_j(W_{\mathcal{I}_L,j}))\leq D_j$.
\end{itemize}
\end{theorem}
\begin{proof}
By Theorem \ref{theorem21}, we can see that each stage can be
realized via a $(2L-1)$-successive Wyner-Ziv scheme.
\end{proof}

The $M$-stage source coding, if realized by concatenating $M$
versions of $(2L-1)$-successive Wyner-Ziv coding schemes, is
essentially a $(2ML-M)$-successive Wyner-Ziv coding scheme. But it
is subject to more restricted conditions since a general
$(2ML-M)$-successive Wyner-Ziv scheme (satisfying the rate
constraints: $R_{\mathcal{I}_L,M}$ and the distortion constraint:
$D_M$) may not be decomposable into $M$ versions of $2L-1$
successive Wyner-Ziv scheme with rate and distortion constraints
satisfied at each stage.

In the remaining part of this section, we shall focus on the
quadratic Gaussian CEO problem.
\begin{lemma}\label{lemma31}
For $R_{\mathcal{I}_L,1}\leq R_{\mathcal{I}_L,2}\leq\cdots\leq
R_{\mathcal{I}_L,M}$ and $D_1\geq D_2\geq\cdots\geq D_M$, the
$M$-stage source coding
\begin{equation*}
(R_{\mathcal{I}_L,1},D_1)\nearrow
(R_{\mathcal{I}_L,2},D_2)\nearrow\cdots\nearrow
(R_{\mathcal{I}_L,M},D_M)
\end{equation*}
is feasible if there exist $r_{\mathcal{I}_L,j}\in\mathbb{R}_+^L$,
$j=1,2,\cdots,M$, satisfying
\begin{itemize}
\item [(i)] $r_{\mathcal{I}_L,j-1}\leq
r_{\mathcal{I}_L,j}$ for all $j\in\mathcal{I}_M$,
\item [(ii)]
$1/\sigma^2_X+\sum_{i=1}^L
\left(1-\exp(-2r_{i,j})\right)/\sigma^2_{N_i}= 1/D_j$ for all
$j\in\mathcal{I}_M$,
\end{itemize}
such that
\begin{eqnarray}
\sum\limits_{i\in\mathcal{A}}(R_{i,j}-R_{i,j-1})&\geq&\frac{1}{2}\log\frac{1}{D_j}-\frac{1}{2}\log\left(\frac{1}{\sigma^2_X}+\sum\limits_{i\in\mathcal{A}}\frac{1-\exp(-2r_{i,j-1})}{\sigma^2_{N_i}}+\sum\limits_{i\in\mathcal{A}^c}\frac{1-\exp(-2r_{i,j})}{\sigma^2_{N_i}}\right)\nonumber\\
&&+\sum\limits_{i\in\mathcal{A}}(r_{i,j}-r_{i,j-1}),\quad \forall
j\in\mathcal{I}_M, \forall \mbox{ nonempty set }
\mathcal{A}\subseteq\mathcal{I}_L.
\end{eqnarray}
Here we assume $r_{\mathcal{I}_L,0}=(0,\cdots,0)$.
\end{lemma}
\begin{proof}
Let $W_{i,M}=Y_i+T_{i,M}$ and
$W_{i,j}=W_{i,j+1}+T_{i,j}\mbox{ }(j\in\mathcal{I}_{M-1})$, where
$T_{i,j}\sim\mathcal{N}(0,\sigma^2_{T_{i,j}}), i\in\mathcal{I}_L,
j\in\mathcal{I}_M $ are all independent and they are also
independent of $(X,Y_{\mathcal{I}_L})$. Let
\begin{equation}
r_{i,j}=I(Y_i;W_{i,j}|X)=\frac{1}{2}\log\frac{\sigma^2_{N_i}+\sum\limits_{k=j}^M\sigma^2_{T_{i,j}}}{\sum\limits_{k=j}^M\sigma^2_{T_{i,j}}}\label{3}
\end{equation}
and $\mathbb{E}(X-\mathbb{E}(X|W_{\mathcal{I}_L,j}))^2=D_j$ for
all $j\in\mathcal{I}_M$, i.e.,
\begin{equation}
\frac{1}{\sigma^2_X}+\sum_{i=1}^L
\frac{1-\exp(-2r_{i,j})}{\sigma^2_{N_i}}=\frac{1}{D_j}.
\end{equation}

Let $R_{\mathcal{I}_L,0}=(0,\cdots,0)$ and let
$W_{\mathcal{I}_L,0}$ be a constant vector. By Theorem
\ref{thm31}, for any $R_{\mathcal{I}_L,1}\leq
R_{\mathcal{I}_L,2}\leq\cdots\leq R_{\mathcal{I}_L,M}$ with
\begin{equation}
\left(R_{\mathcal{I}_L,j}-R_{\mathcal{I}_L,j-1}\right)\in\mathcal{R}(W_{\mathcal{I}_L,j}|W_{\mathcal{I}_L,j-1}),\quad\forall
j\in\mathcal{I}_M, \label{21}
\end{equation}
the $M$-stage source coding
\begin{equation*}
(R_{\mathcal{I}_L,1},D_1)\nearrow
(R_{\mathcal{I}_L,2},D_2)\nearrow\cdots\nearrow
(R_{\mathcal{I}_L,M},D_M)
\end{equation*}
is feasible. We can compute (\ref{21}) explicitly as follows:
\begin{eqnarray}
\sum\limits_{i\in\mathcal{A}}(R_{i,j}-R_{i,j-1})&\geq&
I(Y_{A};W_{\mathcal{A},j}|W_{\mathcal{I}_L\backslash\mathcal{A},j},W_{\mathcal{A},j-1})\\
&=&I(X,Y_{A};W_{\mathcal{A},j}|W_{\mathcal{I}_L\backslash\mathcal{A},j},W_{\mathcal{A},j-1})\\
&=&I(X;W_{\mathcal{A},j}|W_{\mathcal{I}_L\backslash\mathcal{A},j},W_{\mathcal{A},j-1})+I(Y_{A};W_{\mathcal{A},j}|X)-I(Y_{A};W_{\mathcal{A},j-1}|X)\\
&=&h(X|W_{\mathcal{I}_L\backslash\mathcal{A},j},W_{\mathcal{A},j-1})-h(X|W_{\mathcal{I}_L,j})+\sum\limits_{i\in\mathcal{A}}(r_{i,j}-r_{i,j-1})\\
&=&\frac{1}{2}\log\frac{1}{D_j}-\frac{1}{2}\log\left(\frac{1}{\sigma^2_X}+\sum\limits_{i\in\mathcal{A}}\frac{1-\exp(-2r_{i,j-1})}{\sigma^2_{N_i}}+\sum\limits_{i\in\mathcal{I}_L\backslash\mathcal{A}}\frac{1-\exp(-2r_{i,j})}{\sigma^2_{N_i}}\right)\nonumber\\
&&+\sum\limits_{i\in\mathcal{A}}(r_{i,j}-r_{i,j-1}),
\quad\forall\mbox{ nonempty set }
\mathcal{A}\subseteq\mathcal{I}_L.
\end{eqnarray}
The proof is now complete.
\end{proof}

\begin{lemma}[\cite{Oohama}, Lemma 1]\label{lemma32}
\begin{equation}
\frac{1}{n}I(\mathbf{X};f^{(n)}_{\mathcal{I}_L,\mathcal{I}_{j}})\geq\frac{1}{2}\log\frac{\sigma^2_X}{D_j},\quad
\forall j\in\mathcal{I}_M.
\end{equation}
\end{lemma}

The next lemma is a direct application of \cite[Lemma 3.2]{TseCEO}
with $C_i=f^{(n)}_{i,\mathcal{I}_k}\mbox{ }(\forall
i\in\mathcal{A})$ and $C_i=f^{(n)}_{i,\mathcal{I}_{j}}\mbox{
}(\forall i\in\mathcal{I}_L\backslash\mathcal{A})$, where
$f^{(n)}_{i,\mathcal{I}_{j}}$ is the abbreviation of
$(f^{(n)}_{i,1}(\mathbf{Y}_i),\cdots,f^{(n)}_{i,j}(\mathbf{Y}_i))$
.
\begin{lemma}\label{lemma33}
Let
$r_{i,j}=\frac{1}{n}I(\mathbf{Y};f^{(n)}_{i,\mathcal{I}_j}|\mathbf{X}),\mbox{
}\forall i\in\mathcal{I}_L, \forall j\in\mathcal{I}_M$. We have,
for all $0\leq j<k\leq M$,
\begin{equation}
\frac{1}{\sigma^2_X}\exp\left(\frac{2}{n}I(\mathbf{X};f^{n}_{\mathcal{A},\mathcal{I}_k},f^{n}_{\mathcal{I}_L\backslash\mathcal{A},\mathcal{I}_{j}})\right)\leq\frac{1}{\sigma^2_X}+\sum\limits_{i\in\mathcal{A}}\frac{1-\exp(-r_{i,k})}{\sigma^2_{N_i}}+\sum\limits_{i\in\mathcal{I}_L\backslash\mathcal{A}}\frac{1-\exp(-r_{i,j})}{\sigma^2_{N_i}},
\label{37}
\end{equation}
where $f^{(n)}_{i,0}\mbox{ }(i\in\mathcal{I}_L)$ are constant
functions and $r_{\mathcal{I}_L,0}=(0,\cdots,0)$.
\end{lemma}

\begin{lemma}\label{lemma34}
For $R_{\mathcal{I}_L,1}\leq R_{\mathcal{I}_L,2}\leq\cdots\leq
R_{\mathcal{I}_L,M}$ and $D_1\geq D_2\geq\cdots\geq D_M$, if the
$M$-stage source coding
\begin{equation*}
(R_{\mathcal{I}_L,1},D_1)\nearrow
(R_{\mathcal{I}_L,2},D_2)\nearrow\cdots\nearrow
(R_{\mathcal{I}_L,M},D_M)
\end{equation*}
is feasible, then there exist
$r_{\mathcal{I}_L,j}\in\mathbb{R}_+^L,\mbox{ }j=1,2,\cdots,M$,
satisfying
\begin{itemize}
\item [(i)] $r_{\mathcal{I}_L,j-1}\leq
r_{\mathcal{I}_L,j}$ for all $j\in\mathcal{I}_M$,
\item [(ii)]
$1/\sigma^2_X+\sum_{i=1}^L
\left(1-\exp(-2r_{i,j})\right)/\sigma^2_{N_i}\geq 1/D_j$ for all
$j\in\mathcal{I}_M$,
\end{itemize}
such that
\begin{eqnarray}
\sum\limits_{i\in\mathcal{A}}(R_{i,k}-R_{i,j})&\geq&\frac{1}{2}\log\frac{1}{D_k}-\frac{1}{2}\log\left(\frac{1}{\sigma^2_X}+\sum\limits_{i\in\mathcal{A}}\frac{1-\exp(-2r_{i,j})}{\sigma^2_{N_i}}+\sum\limits_{i\in\mathcal{I}_L\backslash\mathcal{A}}\frac{1-\exp(-2r_{i,k})}{\sigma^2_{N_i}}\right)\nonumber\\
&&+\sum\limits_{i\in\mathcal{A}}(r_{i,k}-r_{i,j}),\quad
\forall\mbox{ }0\leq j<k\leq M, \forall \mbox{ nonempty set }
\mathcal{A}\subseteq\mathcal{I}_L.
\end{eqnarray}
Here $r_{\mathcal{I}_L,0}=(0,\cdots,0)$.
\end{lemma}
\begin{proof}
Let $r_{i,j}=I(\mathbf{Y};f^{(n)}_{i,\mathcal{I}_j}|\mathbf{X})/n$
$(\forall i\in\mathcal{I}_L, \forall j\in\mathcal{I}_M)$. It is
clear that $r_{\mathcal{I}_L,j-1}\leq r_{\mathcal{I}_L,j}$ for all
$j\in\mathcal{I}_M$. Moreover, if we let
$\mathcal{A}=\mathcal{I}_L$ in (\ref{37}), we have
\begin{equation}
\frac{1}{\sigma^2_X}+\sum\limits_{i=1}^L\frac{1-\exp(-r_{i,k})}{\sigma^2_{N_i}}\geq\frac{1}{\sigma^2_X}\exp\left(\frac{2}{n}I(\mathbf{X};f^{(n)}_{\mathcal{I}_L,\mathcal{I}_k})\right)\geq\frac{1}{D_k},\quad\forall
\mbox{ }k\in\mathcal{I}_M,
\end{equation}
where the last inequality follows from Lemma \ref{lemma32}.

Furthermore, we have
\begin{eqnarray}
\sum\limits_{i\in\mathcal{A}}(R_{i,k}-R_{i,j})&\geq&\frac{1}{n}\sum\limits_{i\in\mathcal{A}}\sum\limits_{s=j+1}^k
H(f^{(n)}_{i,s})
\geq\frac{1}{n}H(f^{(n)}_{\mathcal{A},s},s=j+1,\cdots,k)\\&\geq&\frac{1}{n}I(\mathbf{Y}_{\mathcal{A}};f^{(n)}_{\mathcal{A},s}, s=j+1,\cdots,k|f^{(n)}_{\mathcal{I}_L\backslash\mathcal{A},\mathcal{I}_k},f^{(n)}_{\mathcal{A},\mathcal{I}_{j}})\\
&=&\frac{1}{n}I(\mathbf{X},\mathbf{Y}_{\mathcal{A}};f^{(n)}_{\mathcal{A},s}, s=j+1,\cdots,k|f^{(n)}_{\mathcal{I}_L\backslash\mathcal{A},\mathcal{I}_k},f^{(n)}_{\mathcal{A},\mathcal{I}_{j}})\\
&=&\frac{1}{n}I(\mathbf{X};f^{(n)}_{\mathcal{A},s}, s=j+1,\cdots,k|f^{(n)}_{\mathcal{I}_L\backslash\mathcal{A},\mathcal{I}_k},f^{(n)}_{\mathcal{A},\mathcal{I}_{j}})\nonumber\\
&&+\frac{1}{n}\sum\limits_{i\in\mathcal{A}}I(\mathbf{Y}_i;f^{(n)}_{i,s},s=j+1,\cdots,k|\mathbf{X},f^{(n)}_{i,\mathcal{I}_{j}})\\
&=&\frac{1}{n}I(\mathbf{X};f^{(n)}_{\mathcal{I}_L,\mathcal{I}_{k}})-\frac{1}{n}I(\mathbf{X};f^{(n)}_{\mathcal{I}_L\backslash\mathcal{A},\mathcal{I}_k},f^{(n)}_{\mathcal{A},\mathcal{I}_{j}})+\sum\limits_{i\in\mathcal{A}}(r_{i,k}-r_{i,j})\\
&\geq&\frac{1}{2}\log\frac{1}{D_k}-\frac{1}{2}\log\left(\frac{1}{\sigma^2_X}+\sum\limits_{i\in\mathcal{A}}\frac{1-\exp(-2r_{i,j})}{\sigma^2_{N_i}}+\sum\limits_{i\in\mathcal{I}_L\backslash\mathcal{A}}\frac{1-\exp(-2r_{i,k})}{\sigma^2_{N_i}}\right)\nonumber\\
&&+\sum\limits_{i\in\mathcal{A}}(r_{i,k}-r_{i,j}), \label{32}
\end{eqnarray}
where (\ref{32}) follows from Lemma \ref{lemma32} and Lemma
\ref{lemma33}. Now the proof is complete.
\end{proof}

\begin{lemma}\label{lemma35}
For any $R_{\mathcal{I}_L}\in\mathbb{R}^L_+$, there exists a
unique $r_{\mathcal{I}_L}\in\mathbb{R}^L_+$ satisfying
\begin{equation}
\frac{1}{\sigma^2_X}+\sum\limits_{i=1}^L\frac{1-\exp(-2r_i)}{\sigma^2_{N_i}}\geq\frac{1}{D^*(R_{\mathcal{I}_L})}\label{128}
\end{equation}
and for any nonempty set $\mathcal{A}\subseteq\mathcal{I}_L$
\begin{eqnarray}
\sum\limits_{i\in\mathcal{A}}R_i&\geq&\frac{1}{2}\log\frac{1}{D^*(R_{\mathcal{I}_L})}-\frac{1}{2}\log\left(\frac{1}{\sigma^2_X}+\sum\limits_{i\in\mathcal{I}_L\backslash\mathcal{A}}\frac{1-\exp(-2r_i)}{\sigma^2_{N_i}}\right)+\sum\limits_{i\in\mathcal{A}}r_i.\label{69}
\end{eqnarray}
Denote this $r_{\mathcal{I}_L}$ by
$r^*_{\mathcal{I}_L}(R_{\mathcal{I}_L})$. We have
\begin{equation}
\frac{1}{\sigma^2_X}+\sum\limits_{i=1}^L\frac{1-\exp(-2r^*_i(R_{\mathcal{I}_L}))}{\sigma^2_{N_i}}=\frac{1}{D^*(R_{\mathcal{I}_L})}\label{78}
\end{equation}
and
\begin{equation}
\sum\limits_{i=1}^LR_i=\frac{1}{2}\log\frac{\sigma^2_X}{D^*(R_{\mathcal{I}_L})}+\sum\limits_{i=1}^Lr^*_i(R_{\mathcal{I}_L}).\label{79}
\end{equation}
\end{lemma}
\begin{proof}
See Appendix.
\end{proof}


Now we are ready to prove the main theorem of this section.
\begin{theorem}\label{thm32}
For $R_{\mathcal{I}_L,1}\leq R_{\mathcal{I}_L,2}\leq\cdots\leq
R_{\mathcal{I}_L,M}$, there exists an $M$-stage distributed
successive refinement scheme from $R_{\mathcal{I}_L,1}$ to
$R_{\mathcal{I}_L,2}$, to $\cdots\cdots$, to $R_{\mathcal{I}_L,M}$
if and only if
\begin{eqnarray}
&&\sum\limits_{i\in\mathcal{A}}(R_{i,j}-R_{i,j-1})\nonumber\\
&\geq&\frac{1}{2}\log\frac{1}{D_j^*(R_{\mathcal{I}_L,j})}-\frac{1}{2}\log\left(\frac{1}{\sigma^2_X}+\sum\limits_{i\in\mathcal{A}}\frac{1-\exp(-2r^*_{i}(R_{\mathcal{I}_L,j-1}))}{\sigma^2_{N_i}}+\sum\limits_{i\in\mathcal{I}_L\backslash\mathcal{A}}\frac{1-\exp(-2r^*_{i}(R_{\mathcal{I}_L,j}))}{\sigma^2_{N_i}}\right)\nonumber\\
&&+\sum\limits_{i\in\mathcal{A}}(r^*_{i}(R_{\mathcal{I}_L,j})-r^*_{i}(R_{\mathcal{I}_L,j-1})),\quad
\forall j\in\mathcal{I}_M, \forall \mbox{ nonempty set }
\mathcal{A}\subseteq\mathcal{I}_L. \label{71}
\end{eqnarray}
Here
$R_{\mathcal{I}_L,0}=r^*_{\mathcal{I}_L}(R_{\mathcal{I}_L},0)=(0,\cdots,0)$.
\end{theorem}
\begin{proof}
Let $D_j=D^*(R_{\mathcal{I}_L,j})\mbox{ }(\forall
j\in\mathcal{I}_M)$ in Lemma \ref{lemma34}. Suppose the vector
sequence $r_{\mathcal{I}_L,j}$ $(j=1,2,\cdots,M)$ satisfies all
the constraints in Lemma \ref{lemma34}. By Lemma \ref{lemma35}, we
must have $1/\sigma^2_X+\sum_{i=1}^L
(1-\exp(-2r_{i,j}))/\sigma^2_{N_i}=1/D^*(R_{\mathcal{I}_L,j})$. So
the constraints in Lemma \ref{lemma34} imply the conditions in
Lemma 3.1. Therefore, the conditions in Lemma \ref{lemma31} are
necessary and sufficient. Furthermore, by Lemma \ref{lemma35}
$r_{\mathcal{I}_L,j}$, if exists, must be equal to
$r^*_{\mathcal{I}_L}(R_{\mathcal{I}_L,j})$. The proof is thus
complete.
\end{proof}
Remark: Applying (\ref{79}) and then (\ref{78}), we get
\begin{eqnarray}
\sum\limits_{i=1}^L(R_{i,j}-R_{i,j-1})&=&\frac{1}{2}\log\frac{\sigma^2_X}{D^*(R_{\mathcal{I}_L,j})}+\sum\limits_{i=1}^Lr^*_i(R_{\mathcal{I}_L,j})
-\frac{1}{2}\log\frac{\sigma^2_X}{D^*(R_{\mathcal{I}_L,j-1})}-\sum\limits_{i=1}^Lr^*_i(R_{\mathcal{I}_L,j-1})\\
&=&\frac{1}{2}\log\frac{\sigma^2_X}{D^*(R_{\mathcal{I}_L,j})}-\frac{1}{2}\log\left(\frac{1}{\sigma^2_X}+\sum\limits_{i=1}^L\frac{1-\exp(-2r^*_{i}(R_{\mathcal{I}_L,j-1}))}{\sigma^2_{N_i}}\right)\nonumber\\
&&+\sum\limits_{i=1}^L\left(r^*_i(R_{\mathcal{I}_L,j})-r^*_i(R_{\mathcal{I}_L,j-1})\right),\quad\forall
j\in\mathcal{I}_M. \label{74}
\end{eqnarray}
Hence in (\ref{71}) the constraints on
$\sum_{i=1}^L(R_{i,j}-R_{i,j-1}),j=1,2,\cdots,M,$ are tight.

The sequential structure of (\ref{71}) leads straightforwardly to
the following result.
\begin{corollary}\label{cor31}
For $R_{\mathcal{I}_L,1}\leq R_{\mathcal{I}_L,2}\leq\cdots\leq
R_{\mathcal{I}_L,M}$, there exists an $M$-stage distributed
successive refinement scheme from $R_{\mathcal{I}_L,1}$ to
$R_{\mathcal{I}_L,2}$, to $\cdots\cdots$, to $R_{\mathcal{I}_L,M}$
if and only if there exist a sequence of 2-stage distributed
successive refinement schemes from $R_{\mathcal{I}_L,j-1}$ to
$R_{\mathcal{I}_L,j}$, $j=1,2,\cdots,M$.
\end{corollary}

Corollary \ref{cor31} shows that for the quadratic Gaussian CEO
problem, we only need to focus on 2-stage distributed successive
refinement.

By (\ref{3}), each monotone increasing vector sequence
$r_{\mathcal{I}_L,j}$ $(j=1,2,\cdots,M)$ is associated with a
unique $\sigma^2_{T_{\mathcal{I}_L,j}}$ $(j=1,2,\cdots,M)$ and
thus a unique $W_{\mathcal{I}_L,j}$ $(j=1,2,\cdots,M)$. We shall
let $W^*_{\mathcal{I}_L}(R_{\mathcal{I}_L,j})$ denote the
$W_{\mathcal{I}_L,j}$ that is associated with
$r^*_{\mathcal{I}_L}(R_{\mathcal{I}_L,j})$ $(j=1,2,\cdots,M)$. Now
we state Theorem \ref{thm32} in the following equivalent form,
which highlights the underlying the geometric structure.
\begin{corollary}
For $R_{\mathcal{I}_L,1}\leq R_{\mathcal{I}_L,2}\leq\cdots\leq
R_{\mathcal{I}_L,M}$, there exists an $M$-stage distributed
successive refinement scheme from $R_{\mathcal{I}_L,1}$ to
$R_{\mathcal{I}_L,2}$, to $\cdots\cdots$, to $R_{\mathcal{I}_L,M}$
if and only if
$(R_{\mathcal{I}_L,j}-R_{\mathcal{I}_L,j-1})\in\mathcal{D}(W^*_{\mathcal{I}_L}(R_{\mathcal{I}_L,j})|W^*_{\mathcal{I}_L}(R_{\mathcal{I}_L,j-1}))$,
where
$\mathcal{D}(W^*_{\mathcal{I}_L}(R_{\mathcal{I}_L,j})|W^*_{\mathcal{I}_L}(R_{\mathcal{I}_L,j-1}))$
is the the dominant face of
$\mathcal{R}(W^*_{\mathcal{I}_L}(R_{\mathcal{I}_L,j})|W^*_{\mathcal{I}_L}(R_{\mathcal{I}_L,j-1}))$,
$\forall j\in\mathcal{I}_M$.
\end{corollary}
\begin{proof}
It is easy to verify that (\ref{71}) is equivalent to
\begin{eqnarray}
\sum\limits_{i\in\mathcal{A}}(R_{i,j}-R_{i,j-1})\geq
I\left(Y_{\mathcal{A}};W^*_{\mathcal{A}}(R_{\mathcal{I}_L,j})|W^*_{\mathcal{A}^c}(R_{\mathcal{I}_L,j}),W^*_{\mathcal{A}}(R_{\mathcal{I}_L,j-1})\right),
\quad \forall j\in\mathcal{I}_M, \forall \mbox{ nonempty set }
\mathcal{A}\subseteq\mathcal{I}_L,
\end{eqnarray}
which, by Definition \ref{def23}, is equivalent to
\begin{equation}
\left(R_{\mathcal{I}_L,j}-R_{\mathcal{I}_L,j-1}\right)\in\mathcal{R}(W^*_{\mathcal{I}_L}(R_{\mathcal{I}_L,j})|W^*_{\mathcal{I}_L}(R_{\mathcal{I}_L,j-1})),\quad
\forall j\in\mathcal{I}_M.
\end{equation}
Furthermore, (\ref{74}) is equivalent to
\begin{equation}
\sum\limits_{i=1}^L(R_{i,j}-R_{i,j-1})=I(Y_{\mathcal{I}_L};W^*_{\mathcal{I}_L}(R_{\mathcal{I}_L,j})|W^*_{\mathcal{I}_L,j-1}(R_{\mathcal{I}_L,j})),\quad\forall
j\in\mathcal{I}_M,
\end{equation}
which means $R_{\mathcal{I}_L,j}-R_{\mathcal{I}_L,j-1}$ is on the
dominant face of
$\mathcal{R}(W^*_{\mathcal{I}_L}(R_{\mathcal{I}_L,j})|W^*_{\mathcal{I}_L}(R_{\mathcal{I}_L,j-1}))$,
$\forall j\in\mathcal{I}_M$.
\end{proof}
Remark: Let $\mathcal{F}_i$ be the lowest dimensional face of
$\mathcal{D}(W^*_{\mathcal{I}_L}(R_{\mathcal{I}_L,j})|W^*_{\mathcal{I}_L}(R_{\mathcal{I}_L,j-1}))$
that contains $R_{\mathcal{I}_L,j}-R_{\mathcal{I}_L,j-1}$. By the
discussion in the preceding section, we can see that this
$M$-stage distributed successive refinement can be realized via an
$(ML+\sum_{j=1}^Mdim(\mathcal{F}_j))$-successive Wyner-Ziv coding
scheme.

Now we proceed to compute
$r^*_{\mathcal{I}_L}(R_{\mathcal{I}_L})$. It is easy to see that
$r^*_{\mathcal{I}_L}(R_{\mathcal{I}_L})$ is the maximizer to the
following optimization problem:
\begin{eqnarray}
\max\limits_{r_{\mathcal{I}_L}\in\mathbb{R}^L_+}\frac{1}{\sigma^2_X}+\sum\limits_{i=1}^L\frac{1-\exp(-2r_i)}{\sigma^2_{N_i}}
\end{eqnarray}
subject to
\begin{eqnarray}
\frac{1}{2}\log\left(\frac{\frac{1}{\sigma^2_X}+\sum\limits_{i=1}^L\frac{1-\exp(-2r_i)}{\sigma^2_{N_i}}}{\frac{1}{\sigma^2_X}+\sum\limits_{i\in\mathcal{A}^c}\frac{1-\exp(-2r_i)}{\sigma^2_{N_i}}}\right)
+\sum\limits_{i\in\mathcal{A}}r_i\leq\sum\limits_{i\in\mathcal{A}}R_i,
\quad\forall\mbox{ nonempty set }\mathcal{A}\subset\mathcal{I}_L,
\end{eqnarray}
and
\begin{equation}
\frac{1}{2}\log\left(\frac{1}{\sigma^2_X}+\sum\limits_{i=1}^L\frac{1-\exp(-2r_i)}{\sigma^2_{N_i}}\right)+\frac{1}{2}\log\sigma^2_X
+\sum\limits_{i=1}^Lr_i=\sum\limits_{i=1}^LR_i.
\end{equation}
which is essentially to find the contra-polymatroid
$\mathcal{R}(r_{\mathcal{I}_L})$ that contains $R_{\mathcal{I}_L}$
and has the minimum achievable distortion
$D(r_{\mathcal{I}_L})=1/{\sigma^2_X}+\sum_{i=1}^L(1-\exp(-2r_i))/{\sigma^2_{N_i}}$.
Another approach is use the Lagrangian formulation in the previous
section. That is, first characterize
$r^*_{\mathcal{I}_L}(R_{\mathcal{I}_L})$ for
$R_{\mathcal{I}_L}\in\partial\mathcal{R}(D)$ via studying the
supporting hyperplanes of $\partial\mathcal{R}(D)$ for fixed $D$.
Then change $D$ to get $r^*_{\mathcal{I}_L}(R_{\mathcal{I}_L})$
for all $R_{\mathcal{I}_L}$. This approach is in general more
cumbersome than the first one. But for small $L$, it is relatively
easy to get the parametric expression of
$r^*_{\mathcal{I}_L}(R_{\mathcal{I}_L})$ via the second approach.

To give a concrete example of the distributed successive
refinement, we choose to study the special case where $L=2$. We
shall adopt the second approach. It is easy to see that
$R_{\mathcal{I}_2}$ is either a vertex of
$\mathcal{R}(r^*_1(R_{\mathcal{I}_2}),r^*_2(R_{\mathcal{I}_2}))$
or an interior point of the dominant face (which is a line
segment) of
$\mathcal{R}(r^*_1(R_{\mathcal{I}_2}),r^*_2(R_{\mathcal{I}_2}))$.
For the first case,
$(r^*_1(R_{\mathcal{I}_2}),r^*_2(R_{\mathcal{I}_2}))$ is
completely determined. For the second case, $R_{\mathcal{I}_2}$
must be on the minimum sum-rate line of
$\partial\mathcal{R}(D^*(R_{\mathcal{I}_2}))$. Hence we only need
to study one supporting line of $\partial\mathcal{R}(D)$, namely,
$R_1+R_2=\min_{(R_1,R_2)\in\partial\mathcal{R}(D)}(R_1+R_2)$,
which has been characterized for all $D$ in \cite{ChenCEO}.

Without loss of generality, we assume
$\sigma^2_{N_1}\leq\sigma^2_{N_2}$. Let
\begin{equation}
L_D=\max\left\{k\in\mathcal{I}_2:\frac{k}{\sigma^2_{N_k}}+\frac{1}{D}-\frac{1}{D_{\min}(k)}\geq
0 \right\},
\end{equation}
where
\begin{equation}
\frac{1}{D_{\min}(k)}=\frac{1}{\sigma^2_X}+\sum\limits_{i=1}^k\frac{1}{\sigma^2_{N_i}}.
\end{equation}
Let $\widetilde{D}$ be the unique solution to the following
equation:
\begin{eqnarray}
\frac{1}{2}\log\left(\frac{\sigma^2_X}{D}\prod\limits_{i=1}^{L_D}\left(\frac{L_D}{\sigma^2_{N_i}\left(\frac{1}{D_{\min}(L_D)}-\frac{1}{D}\right)}\right)\right)=R_1+R_2.
\end{eqnarray}
Let
\begin{eqnarray}
\widetilde{r}_1&=&\frac{1}{2}\log\left(\frac{L_{\widetilde{D}}}{\sigma^2_{N_1}\left(\frac{1}{D_{\min}(L_{\widetilde{D}})}-\frac{1}{\widetilde{D}}\right)}\right),\\
\widetilde{r}_2&=&
  \begin{cases}
    0, & L_{\widetilde{D}}=1, \\
    \frac{1}{2}\log\left(\frac{2}{\sigma^2_{N_2}}\left(\frac{1}{\sigma^2_X}+\frac{1}{\sigma^2_{N_1}}+\frac{1}{\sigma^2_{N_2}}-\frac{1}{\widetilde{D}}\right)^{-1}\right), & L_{\widetilde{D}}=2.
  \end{cases}
\end{eqnarray}
We have
\begin{itemize}
\item [(i)] If
\begin{equation}
R_1\geq\frac{1}{2}\log\left(\frac{1}{\sigma^2_X}+\frac{1-\exp(-2\widetilde{r}_1)}{\sigma^2_{N_1}}\right)+\frac{1}{2}\log\sigma^2_X
+\widetilde{r}_1,\label{101}
\end{equation}
then
\begin{eqnarray}
&&\frac{1}{2}\log\left(\frac{1}{\sigma^2_X}+\frac{1-\exp(-2r^*_1(R_{\mathcal{I}_2}))}{\sigma^2_{N_1}}\right)+\frac{1}{2}\log\sigma^2_X
+r^*_1(R_{\mathcal{I}_2})=R_1,\label{102}\\
&&\frac{1}{2}\log\left(1+\sigma^2_X\sum\limits_{i=1}^2\frac{1-\exp(-2r^*_i(R_{\mathcal{I}_2}))}{\sigma^2_{N_i}}\right)
+r^*_1(R_{\mathcal{I}_2})+r^*_2(R_{\mathcal{I}_2})=R_1+R_2.\label{103}
\end{eqnarray}
\item [(ii)] If
\begin{equation}
R_2\geq\frac{1}{2}\log\left(\frac{1}{\sigma^2_X}+\frac{1-\exp(-2\widetilde{r}_2)}{\sigma^2_{N_1}}\right)+\frac{1}{2}\log\sigma^2_X
+\widetilde{r}_2,\label{104}
\end{equation}
then
\begin{eqnarray}
&&\frac{1}{2}\log\left(\frac{1}{\sigma^2_X}+\frac{1-\exp(-2r^*_2(R_{\mathcal{I}_2}))}{\sigma^2_{N_2}}\right)+\frac{1}{2}\log\sigma^2_X
+r^*_2(R_{\mathcal{I}_2})=R_2,\label{105}\\
&&\frac{1}{2}\log\left(1+\sigma^2_X\sum\limits_{i=1}^2\frac{1-\exp(-2r^*_i(R_{\mathcal{I}_2}))}{\sigma^2_{N_i}}\right)
+r^*_1(R_{\mathcal{I}_2})+r^*_2(R_{\mathcal{I}_2})=R_1+R_2.\label{106}
\end{eqnarray}
\item [(iii)] Otherwise
$r^*_i(R_{\mathcal{I}_2})=\widetilde{r}_i, i=1,2.$
\end{itemize}
The above three conditions essentially divide $\mathbb{R}^2_+$
into 3 regions. Define
\begin{eqnarray}
\Omega_1&=&\left\{(R_1,R_2)\in\mathbb{R}^2_+:
R_1\geq\frac{1}{2}\log\left(\frac{1}{\sigma^2_X}+\frac{1-\exp(-2\widetilde{r}_1)}{\sigma^2_{N_1}}\right)+\frac{1}{2}\log\sigma^2_X
+\widetilde{r}_1\right\},\\
\Omega_3&=&\left\{(R_1,R_2)\in\mathbb{R}^2_+:
R_2\geq\frac{1}{2}\log\left(\frac{1}{\sigma^2_X}+\frac{1-\exp(-2\widetilde{r}_2)}{\sigma^2_{N_2}}\right)+\frac{1}{2}\log\sigma^2_X
+\widetilde{r}_2\right\},\\
\Omega_3&=&\left\{(R_1,R_2)\in\mathbb{R}^2_+:
R_i\leq\frac{1}{2}\log\left(\frac{1}{\sigma^2_X}+\frac{1-\exp(-2\widetilde{r}_i)}{\sigma^2_{N_i}}\right)+\frac{1}{2}\log\sigma^2_X
+\widetilde{r}_i, i=1,2\right\}.
\end{eqnarray}
It is easy to check that except the boundaries (i.e., those rate
tuples that satisfy (\ref{101}) or (\ref{101}) with equality),
$\Omega_1,\Omega_2$ and $\Omega_3$ do not overlap. Typical shapes
of $\Omega_1,\Omega_2$ and $\Omega_3$ are plotted in Fig. 2. Any
rate pair $R_{\mathcal{I}_2}\in\{\Omega_1\cup\Omega_2\}$  is a
vertex of
$\mathcal{R}(r^*_1(R_{\mathcal{I}_2}),r^*_2(R_{\mathcal{I}_2}))$
and thus is associated with a 2-successive Wyner-Ziv coding
scheme. Any rate pair $R_{\mathcal{I}_2}$ strictly inside
$\Omega_3$ is an interior point of the dominant face of
$\mathcal{R}(r^*_1(R_{\mathcal{I}_2}),r^*_2(R_{\mathcal{I}_2}))$
and thus is associated with a 3-successive Wyenr-Ziv coding
scheme. Hence there is a clear distinction between
$(\Omega_1,\Omega_2)$ and $\Omega_3$. We will see that this
difference manifests itself in the behavior of distributed
successive refinement.

Henceforth we shall assume $R_{\mathcal{I}_2,2}\geq
R_{\mathcal{I}_2,1}$.
\begin{claim}\label{claim31}
$(r^*_1(R_{\mathcal{I}_2,2}),r^*_2(R_{\mathcal{I}_2,2}))\geq
(r^*_1(R_{\mathcal{I}_2,1}),r^*_2(R_{\mathcal{I}_2,1}))$.
\end{claim}
\begin{proof}
If both $R_{\mathcal{I}_2,1}$ and $R_{\mathcal{I}_2,2}$ are in
$\Omega_1$ or both $R_{\mathcal{I}_2,1}$ and $R_{\mathcal{I}_2,2}$
are in $\Omega_2$, the claim can be easily verified by checking
the equations (\ref{102}), (\ref{103}), (\ref{105}) and
(\ref{106}). Since $\widetilde{r}_1$ and $\widetilde{r}_2$ are
monotone increasing functions of $R_1+R_2$, the claim is also true
when both $R_{\mathcal{I}_2,1}$ and $R_{\mathcal{I}_2,2}$ are in
$\Omega_3$.

Now consider the general case when $R_{\mathcal{I}_2,1}$ and
$R_{\mathcal{I}_2,2}$ are in different regions, say
$R_{\mathcal{I}_2,1}\in\Omega_1$ and
$R_{\mathcal{I}_2,2}\in\Omega_3$. Suppose the line segment that
connects $R_{\mathcal{I}_2,1}$ and $R_{\mathcal{I}_2,2}$
intersects the boundary of $\Omega_1$ and $\Omega_3$ at point
$R'_{\mathcal{I}_2}$. We have
$(r^*_1(R'_{\mathcal{I}_2}),r^*_2(R'_{\mathcal{I}_2}))\geq
(r^*_1(R_{\mathcal{I}_2,1}),r^*_2(R_{\mathcal{I}_2,1}))$ since
both $R_{\mathcal{I}_2,1}$ and $R'_{\mathcal{I}_2}$ are in
$\Omega_1$ and
$(r^*_1(R_{\mathcal{I}_2,2}),r^*_2(R_{\mathcal{I}_2,2}))\geq
(r^*_1(R'_{\mathcal{I}_2}),r^*_2(R'_{\mathcal{I}_2}))$ since both
$R'_{\mathcal{I}_2}$ and $R_{\mathcal{I}_2,2}$ are in $\Omega_3$.
Hence $(r^*_1(R_{\mathcal{I}_2,2}),r^*_2(R_{\mathcal{I}_2,2}))\geq
(r^*_1(R_{\mathcal{I}_2,1}),r^*_2(R_{\mathcal{I}_2,1}))$. The
other cases can be discussed in a similar way.
\end{proof}

\begin{claim}\label{claim32}
If both $R_{\mathcal{I}_2,1}$ and $R_{\mathcal{I}_2,2}$ are in
$\Omega_1$, then there exists a distributed successive refinement
scheme from $R_{\mathcal{I}_2,1}$ to $R_{\mathcal{I}_2,2}$ if and
only if $R_{1,2}=R_{1,1}$ or $R_{2,1}=0$.
\end{claim}
\begin{proof}
If $R_{1,2}=R_{1,1}$, by (\ref{102}) we have
$r^*_1(R_{\mathcal{I}_2,2})=r^*_1(R_{\mathcal{I}_2,1})$. It is
easy to verify that the conditions in Theorem \ref{thm32} are all
satisfied. If $R_{2,1}=0$, by (\ref{102}) and (\ref{103}), we have
$r^*_2(R_{\mathcal{I}_2,1})=0$. Again, it is easy to check that
the conditions in Theorem \ref{thm32} are all satisfied.

Now suppose there exists a distributed successive refinement
scheme from $R_{\mathcal{I}_2,1}$ to $R_{\mathcal{I}_2,2}$.  Since
both $R_{\mathcal{I}_2,1}$ and $R_{\mathcal{I}_2,2}$ are in
$\Omega_1$, by (\ref{102}) and (\ref{103})
\begin{eqnarray*}
&&R_{2,2}-R_{2,1}\\
&=&\frac{1}{2}\log\left(\frac{1}{\sigma^2_X}+\sum\limits_{i=1}^2\frac{1-\exp(-2r^*_i(R_{\mathcal{I}_2,2}))}{\sigma^2_{N_i}}\right)+r^*_{2}(R_{\mathcal{I}_2,2})-\frac{1}{2}\log\left(\frac{1}{\sigma^2_X}+\frac{1-\exp(-2r^*_1(R_{\mathcal{I}_2,2}))}{\sigma^2_{N_1}}\right)\\
&&-\frac{1}{2}\log\left(\frac{1}{\sigma^2_X}+\sum\limits_{i=1}^2\frac{1-\exp(-2r^*_i(R_{\mathcal{I}_2,1}))}{\sigma^2_{N_i}}\right)-r^*_{2}(R_{\mathcal{I}_2,1})+\frac{1}{2}\log\left(\frac{1}{\sigma^2_X}+\frac{1-\exp(-2r^*_1(R_{\mathcal{I}_2,1}))}{\sigma^2_{N_1}}\right)\\
&=&\frac{1}{2}\log\frac{1}{D^*(R_{\mathcal{I}_2,2})}-\frac{1}{2}\log\left(\frac{1}{\sigma^2_X}+\frac{1-\exp(-2r^*_1(R_{\mathcal{I}_2,2}))}{\sigma^2_{N_1}}\right)+r^*_{2}(R_{\mathcal{I}_2,2})-r^*_{2}(R_{\mathcal{I}_2,1})
\\&&-\frac{1}{2}\log\left(\frac{1}{\sigma^2_X}+\sum\limits_{i=1}^2\frac{1-\exp(-2r^*_i(R_{\mathcal{I}_2,1}))}{\sigma^2_{N_i}}\right)
+\frac{1}{2}\log\left(\frac{1}{\sigma^2_X}+\frac{1-\exp(-2r^*_1(R_{\mathcal{I}_2,1}))}{\sigma^2_{N_1}}\right).
\end{eqnarray*}
By Theorem \ref{thm32}, we must have
\begin{eqnarray*}
&&\frac{1}{2}\log\frac{1}{D^*(R_{\mathcal{I}_2,2})}-\frac{1}{2}\log\left(\frac{1}{\sigma^2_X}+\frac{1-\exp(-2r^*_1(R_{\mathcal{I}_2,2}))}{\sigma^2_{N_1}}\right)+r^*_{2}(R_{\mathcal{I}_2,2})-r^*_{2}(R_{\mathcal{I}_2,1})
\\&&-\frac{1}{2}\log\left(\frac{1}{\sigma^2_X}+\sum\limits_{i=1}^2\frac{1-\exp(-2r^*_i(R_{\mathcal{I}_2,1}))}{\sigma^2_{N_i}}\right)
+\frac{1}{2}\log\left(\frac{1}{\sigma^2_X}+\frac{1-\exp(-2r^*_1(R_{\mathcal{I}_2,1}))}{\sigma^2_{N_1}}\right)\\
&\geq&\frac{1}{2}\log\frac{1}{D^*(R_{\mathcal{I}_2,2})}-\frac{1}{2}\log\left(\frac{1}{\sigma^2_X}+\frac{1-\exp(-2r^*_1(R_{\mathcal{I}_2,2}))}{\sigma^2_{N_1}}+\frac{1-\exp(-2r^*_2(R_{\mathcal{I}_2,1}))}{\sigma^2_{N_2}}\right)\\
&&+r^*_{2}(R_{\mathcal{I}_2,2})-r^*_{2}(R_{\mathcal{I}_2,1}),
\end{eqnarray*}
which, after some algebraic manipulation, is equivalent to
$r^*_1(R_{\mathcal{I}_2,2})r^*_2(R_{\mathcal{I}_2,1})\leq
r^*_1(R_{\mathcal{I}_2,1})r^*_2(R_{\mathcal{I}_2,1})$. Then we
have either $r^*_1(R_{\mathcal{I}_2,2})\leq
r^*_1(R_{\mathcal{I}_2,1})$ (which further implies
$r^*_1(R_{\mathcal{I}_2,2})=r^*_1(R_{\mathcal{I}_2,1})$) or
$r^*_2(R_{\mathcal{I}_2,1})=0$ . Hence, by (\ref{102}) and
(\ref{103}), we have $R_{1,2}=R_{1,1}$ or $R_{2,1}=0$.
\end{proof}

The following claim follows by symmetry.
\begin{claim}\label{claim33}
If both $R_{\mathcal{I}_2,1}$ and $R_{\mathcal{I}_2,2}$ are in
$\Omega_2$, then there exists a distributed successive refinement
scheme from $R_{\mathcal{I}_2,1}$ to $R_{\mathcal{I}_2,2}$ if and
only if $R_{2,2}=R_{2,1}$ or $R_{1,1}=0$.
\end{claim}
Remark: Claim \ref{claim32} and \ref{claim33} imply that there
exists a distributed successive refinement scheme from
$R_{\mathcal{I}_2,1}$ to $R_{\mathcal{I}_2,2}$ if
$R_{\mathcal{I}_2,1}$ and $R_{\mathcal{I}_2,2}$ are on the
$R_1$-axis or $R_{\mathcal{I}_2,1}$ and $R_{\mathcal{I}_2,2}$ are
on the $R_2$-axis. Actually in this case, the distributed
successive refinement reduces to the conventional successive
refinement in the single source coding\footnote{There is a slight
difference since the CEO problem, after reduced to the single
encoder case, becomes the noisy (single) source coding problem.
But the generalization of the successive refinement in the single
source coding to the noisy (single) source coding is
straightforward.} \cite{Equitz}. Furthermore, if $R_1=\infty$ and
$\sigma^2_{N_2}=0$ (or $R_2=\infty$ and $\sigma^2_{N_1}=0$), then
the quadratic Gaussian CEO problem becomes the Wyner-Ziv problem
of jointly Gaussian source. Claim \ref{claim32} (or Claim
\ref{claim33}) implies the successive refinability for the
Wyner-Ziv problem of jointly Gaussian sources \cite{Steinberg}.

\begin{claim}
Suppose $R_{1,1}>0, R_{2,1}>0$. Then there is no distributed
successive refinement scheme from $R_{\mathcal{I}_2,1}$ to
$R_{\mathcal{I}_2,2}$ if $R_{\mathcal{I}_2,1}\in\Omega_1$,
$R_{\mathcal{I}_2,2}\in\Omega_2$ or
$R_{\mathcal{I}_2,1}\in\Omega_2$,
$R_{\mathcal{I}_2,2}\in\Omega_1$.
\end{claim}
\begin{proof}
We shall only prove the case for $R_{\mathcal{I}_2,1}\in\Omega_1$,
$R_{\mathcal{I}_2,2}\in\Omega_2$. The other one follows by
symmetry.

By (\ref{102}) and (\ref{103}), $R_{1,1}>0, R_{2,1}>0$ implies
$r^*_1(R_{\mathcal{I}_2,1})>0, r^*_2(R_{\mathcal{I}_2,1})>0$,
which further implies $r^*_2(R_{\mathcal{I}_2,2})>0$ by Claim
\ref{claim31}. Now it follows from (\ref{102}), (\ref{103}),
(\ref{105}) and (\ref{106}) that
\begin{eqnarray*}
&&R_{1,2}-R_{1,1}\\
&=&\frac{1}{2}\log\left(\frac{1}{\sigma^2_X}+\sum\limits_{i=1}^2\frac{1-\exp(-2r^*_i(R_{\mathcal{I}_2,2}))}{\sigma^2_{N_i}}\right)+r^*_{1}(R_{\mathcal{I}_2,2})-\frac{1}{2}\log\left(\frac{1}{\sigma^2_X}+\frac{1-\exp(-2r^*_2(R_{\mathcal{I}_2,2}))}{\sigma^2_{N_2}}\right)\\
&&-\frac{1}{2}\log\left(\frac{1}{\sigma^2_X}+\frac{1-\exp(-2r^*_1(R_{\mathcal{I}_2,1}))}{\sigma^2_{N_1}}\right)-\frac{1}{2}\log\sigma^2_X
-r^*_1(R_{\mathcal{I}_2,1})\\
&=&\frac{1}{2}\log\frac{1}{D^*(R_{\mathcal{I}_2,2})}
-\frac{1}{2}\log\left(\frac{1}{\sigma^2_X}+\frac{1-\exp(-2r^*_2(R_{\mathcal{I}_2,2}))}{\sigma^2_{N_2}}\right)
-\frac{1}{2}\log\sigma^2_X\\
&&-\frac{1}{2}\log\left(\frac{1}{\sigma^2_X}+\frac{1-\exp(-2r^*_1(R_{\mathcal{I}_2,1}))}{\sigma^2_{N_1}}\right)
+r^*_{1}(R_{\mathcal{I}_2,2})-r^*_1(R_{\mathcal{I}_2,1}),
\end{eqnarray*}
which is strictly less than
\begin{eqnarray*}
\frac{1}{2}\log\frac{1}{D^*(R_{\mathcal{I}_2,2})}-\frac{1}{2}\log\left(\frac{1}{\sigma^2_X}+\frac{1-\exp(-2r^*_2(R_{\mathcal{I}_2,2}))}{\sigma^2_{N_2}}+\frac{1-\exp(-2r^*_1(R_{\mathcal{I}_2,1}))}{\sigma^2_{N_1}}\right)
+r^*_{1}(R_{\mathcal{I}_2,2})-r^*_1(R_{\mathcal{I}_2,1})
\end{eqnarray*}
if $r^*_1(R_{\mathcal{I}_2,1})>0, r^*_2(R_{\mathcal{I}_2,2})>0$.
Thus by Theorem \ref{thm32}, the distributed successive refinement
scheme can not exist.
\end{proof}

\begin{figure}[hbt]\label{fig2}
\centering
\includegraphics[scale=0.5]{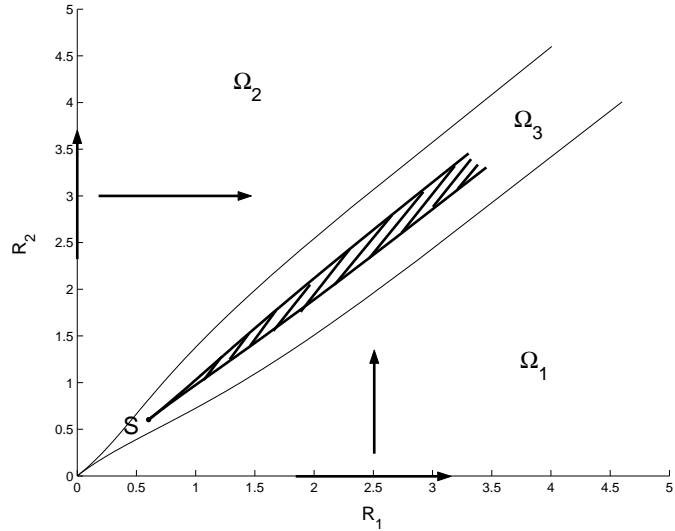}
\caption{Distributed successive refinement for the quadratic Gaussian CEO problem}
\end{figure}

In Fig. 2, the arrows denote the possible directions for the
distributed successive refinement in $\Omega_1$ and $\Omega_3$.
For illustration, we pick a point $s$ in $\Omega_2$. The dark
region is the set of points to which there exists a distributed
successive refinement scheme from $s$. We can see that the
distributed successive refinement behaves very differently in
these three regions. 

\section{Conclusion}
We discussed two closely related problems in distributed source
coding: The first one is how to decompose a high complexity
distributed source code into low complexity codes; The second one
is how to construct a high rate distributed source code using low
rate codes via distributed successive refinement. It turns out
that, at least for the quadratic Gaussian CEO problem, the
successive Wyner-Ziv coding scheme gives the answer to both
problems. Besides the features (say, low complexity and
robustness) we discussed in the paper, the concatenable chain
structure of the successive Wyner-Ziv coding scheme seems
especially attractive in wireless sensor networks, where channels
are subject to fluctuation. In this case, by properly converting a
high-rate distributed source code to a multistage code via the
successive Wyner-Ziv coding scheme, one can match source rates to
the channel rates adaptively.

\appendix[Proof of Lemma 3.5]

For any $r_{\mathcal{I}_L}\in\mathbb{R}^L_+$, define two set
functions $f(\cdot,r_{\mathcal{I}_L}),
f_D(\cdot,r_{\mathcal{I}_L}):2^{\mathcal{I}_L}\rightarrow
\mathbb{R}_+$:
\begin{eqnarray*}
f(\mathcal{A},r_{\mathcal{I}_L})&=&\frac{1}{2}\log\left(\frac{\frac{1}{\sigma^2_X}+\sum\limits_{i=1}^L\frac{1-\exp(-2r_i)}{\sigma^2_{N_i}}}{\frac{1}{\sigma^2_X}+\sum\limits_{i\in\mathcal{I}_L\backslash\mathcal{A}}\frac{1-\exp(-2r_i)}{\sigma^2_{N_i}}}\right)
+\sum\limits_{i\in\mathcal{A}}r_i, \forall\mbox{ nonempty set
}\mathcal{A}\subseteq\mathcal{I}_L,\\
f_D(\mathcal{A},r_{\mathcal{I}_L})&=&\frac{1}{2}\log\frac{1}{D}-\frac{1}{2}\log\left(\frac{1}{\sigma^2_X}+\sum\limits_{i\in\mathcal{I}_L\backslash\mathcal{A}}\frac{1-\exp(-2r_i)}{\sigma^2_{N_i}}\right)
+\sum\limits_{i\in\mathcal{A}}r_i, \forall\mbox{ nonempty set
}\mathcal{A}\subseteq\mathcal{I}_L,
\end{eqnarray*}
and
$f(\emptyset,r_{\mathcal{I}_L})=f_D(\emptyset,r_{\mathcal{I}_L})=0$.

Note that $f(\cdot,r_{\mathcal{I}_L})$ is a rank function and
induces the contra-polymatroid $\mathcal{R}(r_{\mathcal{I}_L})$
defined in (\ref{17}). Furthermore, for any nonempty set
$\mathcal{A}\subseteq\mathcal{I}_L$,
\begin{eqnarray}
f(\mathcal{A},r_{\mathcal{I}_L})-f_D(\mathcal{A},r_{\mathcal{I}_L})&=&\frac{1}{2}\log\left(\frac{1}{\sigma^2_X}+\sum\limits_{i=1}^L\frac{1-\exp(-2r_i)}{\sigma^2_{N_i}}\right)-\frac{1}{2}\log\frac{1}{D}.\label{222}
\end{eqnarray}

By the supermodular property of $f(\cdot,r_{\mathcal{I}_L})$ and
the equation (\ref{222}), we can establish that, for any
$r_{\mathcal{I}_L}$ satisfying $r_i>0$ $(\forall
i\in\mathcal{I}_L)$ and nonempty sets
$\mathcal{S},\mathcal{T}\subseteq\mathcal{I}_L$,
\begin{itemize}
\item [(i)]
\begin{equation}
f(\mathcal{S},r_{\mathcal{I}_L})+f(\mathcal{T},r_{\mathcal{I}_L})<
f(\mathcal{S}\cup\mathcal{T},r_{\mathcal{I}_L})+f(\mathcal{S}\cap\mathcal{T},r_{\mathcal{I}_L});
\label{41}
\end{equation}
\item [(ii)] If $1/\sigma^2_X+\sum_{i=1}^L
(1-\exp(-2r_{i}))/\sigma^2_{N_i}\geq 1/D$, and
$\mathcal{S}\nsubseteq\mathcal{T}$,
$\mathcal{T}\nsubseteq\mathcal{S}$, then
\begin{equation}
f_D(\mathcal{S},r_{\mathcal{I}_L})+f_D(\mathcal{T},r_{\mathcal{I}_L})<f_D(\mathcal{S}\cup\mathcal{T},r_{\mathcal{I}_L})+f_D(\mathcal{S}\cap\mathcal{T},r_{\mathcal{I}_L}).\label{333}
\end{equation}
\end{itemize}

It was shown in \cite{TseCEO} that
\begin{equation}
\mathcal{R}(D)=\bigcup\limits_{r_{\mathcal{I}_L}\in\mathcal{F}(D)}\left\{R_{\mathcal{I}_L}:\sum\limits_{i\in\mathcal{A}}R_i\geq
f_D(\mathcal{A},r_{\mathcal{I}_L}), \forall\mbox{ nonempty set
}\mathcal{A}\subseteq\mathcal{I}_L\right\}, \label{vinod}
\end{equation}
where $\mathcal{F}(D)$ is defined in (\ref{8}). Hence there must
exist a vector $r_{\mathcal{I}_L}\in\mathbb{R}^L_+$ satisfying the
constraints (\ref{128}) and (\ref{69}) in Lemma \ref{lemma35},
i.e.,
\begin{eqnarray}
\sum\limits_{i\in\mathcal{A}}R_{i}\geq
f_{D^*(R_{\mathcal{I}_L})}(\mathcal{A},r_{\mathcal{I}_L}),\quad\forall\mbox{
nonempty set }\mathcal{A}\subseteq\mathcal{I}_L,\label{52}
\end{eqnarray}
and
\begin{equation}
\frac{1}{\sigma^2_X}+\sum\limits_{i=1}^L\frac{1-\exp(-r_{i})}{\sigma^2_{N_i}}\geq\frac{1}{D^*(R_{\mathcal{I}_L})}.\label{53}
\end{equation}
Let $\mathcal{G}=\{i\in\mathcal{I}_L: r_{i}>0\}$. Then (\ref{52})
and (\ref{53}) reduce to the following constraints:
\begin{eqnarray}
\sum\limits_{i\in\mathcal{A}}R_{i}\geq
f_{D^*(R_{\mathcal{I}_L})}(\mathcal{A},r_{\mathcal{I}_L}),\quad\forall\mbox{
nonempty set }\mathcal{A}\subseteq\mathcal{G},
\end{eqnarray}
and
\begin{equation}
\frac{1}{\sigma^2_X}+\sum\limits_{i\in\mathcal{G}}\frac{1-\exp(-r_{i})}{\sigma^2_{N_i}}\geq\frac{1}{D^*(R_{\mathcal{I}_L})}.
\end{equation}
are still active. Thus without loss of generality, we can assume
$\mathcal{G}=\mathcal{I}_L$.

It can be shown that in (\ref{52}), if the constraints on
$\sum_{i\in\mathcal{S}}R_i$ and $\sum_{i\in\mathcal{T}}R_i$ are
tight, then either $\mathcal{S}\subseteq\mathcal{T}$ or
$\mathcal{T}\subseteq\mathcal{S}$. Otherwise
\begin{eqnarray}
&&f_{D^*(R_{\mathcal{I}_L})}(\mathcal{S},r_{\mathcal{I}_L})+f_{D^*(R_{\mathcal{I}_L})}(\mathcal{T},r_{\mathcal{I}_L})
\\&=&\sum\limits_{i\in\mathcal{S}}R_{i}+\sum\limits_{i\in\mathcal{T}}R_{i}\\
&=&\sum\limits_{i\in\mathcal{S}\cup\mathcal{T}}R_{i,j}+\sum\limits_{i\in\mathcal{S}\cap\mathcal{T}}R_{i}\\
&\geq&f_{D^*(R_{\mathcal{I}_L})}(\mathcal{S}\cup\mathcal{T},r_{\mathcal{I}_L})+f_{D^*(R_{\mathcal{I}_L})}(\mathcal{S}\cap\mathcal{T},r_{\mathcal{I}_L}),
\end{eqnarray}
contradictory to (\ref{333}). Let
$\mathcal{\widetilde{A}}=\bigcap_{k\in\mathcal{I}_K}\mathcal{A}_k$,
where $\mathcal{A}_k (k\in\mathcal{I}_K)$ are the sets for which
the constraints on $\sum_{i\in\mathcal{A}_k}R_i$ are tight in
(\ref{52}). If there is no such an $\mathcal{A}_k$, let
$\mathcal{\widetilde{A}}=\mathcal{I}_L$. $\mathcal{\widetilde{A}}$
is thus always nonempty.

Now suppose
\begin{equation}
\frac{1}{\sigma^2_X}+\sum\limits_{i=1}^L\frac{1-\exp(-r_{i})}{\sigma^2_{N_i}}>\frac{1}{D^*(R_{\mathcal{I}_L})}.\label{54}
\end{equation}
Pick any $i^*\in\widetilde{A}$, we can decreases $r_{i^*}$ to
$r_{i^*}-\delta$ for some $\delta>0$ so that all the constraints
in (\ref{52}) and (\ref{54}) become non-tight.  Then we can
decrease $D^*(R_{\mathcal{I}_L})$ to
$D^*(R_{\mathcal{I}_L})-\epsilon$ for some $\epsilon>0$ without
violating any constraints in (\ref{52}) and (\ref{54}). By
(\ref{vinod}) we have
$R_{\mathcal{I}_L}\in\mathcal{R}(D^*(R_{\mathcal{I}_L})-\epsilon)$,
contradictory to the definition of $D^*(R_{\mathcal{I}_L})$. Hence
we can conclude that (\ref{78}) holds, i.e.,
\begin{equation}
\frac{1}{\sigma^2_X}+\sum\limits_{i=1}^L\frac{1-\exp(-r_{i})}{\sigma^2_{N_i}}=\frac{1}{D^*(R_{\mathcal{I}_L})}.\label{equal}
\end{equation}

Now we proceed to show that $r_{\mathcal{I}_L}$ must be unique.

It is easy to check that $1/\sigma^2_X+\sum_{i=1}^L
(1-\exp(-2r_{i}))/\sigma^2_{N_i}$ is a strict concave function of
$r_{\mathcal{I}_L}$ and for any nonempty set
$\mathcal{A}\subseteq\mathcal{I}_L$,
$f_D(\mathcal{A},r_{\mathcal{I}_L})$ is convex in
$r_{\mathcal{I}_L}$.

Suppose both $r'_{\mathcal{I}_L}$ and
$r''_{\mathcal{I}_L}\in\mathbb{R}^L_+$ satisfy the constraints
(\ref{52}) and (\ref{53}), and there exists some $i^*$ such that
$r'_{i^*}\neq r''_{i^*}$. We shall first show that $r'_{i^*},
r''_{i^*}$ are both finite. If not, without loss of generality
suppose $r'_{i^*}=\infty$, which implies that $R_{i^*}=\infty$.
Now construct a new vector $r'''_{\mathcal{I}_L}$ such that
$r'''_{i}=r'_{i}=\infty$ if $i=i^*$ and $r'''_{i}=r''_{i}$
otherwise. Note: we have $r'''_{i^*}>r''_{i^*}$. It is easy to
check that $r'''_{\mathcal{I}_L}$ satisfies the constraints
(\ref{52}) and (\ref{53}) (Note: we let $\infty-\infty=0$). But we
have
\begin{equation}
\frac{1}{\sigma^2_X}+\sum\limits_{i=1}^L\frac{1-\exp(-2r'''_{i})}{\sigma^2_{N_i}}>\frac{1}{\sigma^2_X}+\sum\limits_{i=1}^L\frac{1-\exp(-2r''_{i})}{\sigma^2_{N_i}}=\frac{1}{D^*(R_{\mathcal{I}_L})},
\end{equation}
which is contradictory to (\ref{equal}).

Now let $\overline{r}_{i}=(r'_{i}+r''_{i})/2$ for all
$i\in\mathcal{I}_L$. Note that $\overline{r}_{i^*}$ is equal to
neither $r'_{i^*}$ nor $r''_{i^*}$ since $r'_{i^*}\neq r''_{i^*}$
and both are finite. It is obvious that
$\overline{r}_{\mathcal{I}_L}\in\mathbb{R}_+^L$. Furthermore, we
have
\begin{eqnarray}
\frac{1}{\sigma^2_X}+\sum\limits_{i=1}^L\frac{1-\exp(-\overline{r}_{i})}{\sigma^2_{N_i}}&\geq&\frac{1}{\sigma^2_X}+\frac{1}{2}\sum\limits_{i=1}^L\frac{1-\exp(-r'_{i})}{\sigma^2_{N_i}}+\frac{1}{2}\sum\limits_{i=1}^L\frac{1-\exp(-r''_{i})}{\sigma^2_{N}}\label{63}
\\&\geq&\frac{1}{D^*(R_{\mathcal{I}_L})},
\end{eqnarray}
and
\begin{eqnarray}
\sum\limits_{i\in\mathcal{A}}R_i&\geq&\frac{1}{2}f_{D^*(R_{\mathcal{I}_L})}(\mathcal{A},r'_{\mathcal{I}_L})+\frac{1}{2}f_{D^*(R_{\mathcal{I}_L})}(\mathcal{A},r''_{\mathcal{I}_L})\\
&\geq&f_{D^*(R_{\mathcal{I}_L})}(\mathcal{A},\overline{r}_{\mathcal{I}_L}),\quad\forall
\mbox{ nonempty set } \mathcal{A}\subseteq\mathcal{I}_L.
\end{eqnarray}
Hence $\overline{r}_{\mathcal{I}_L}$ satisfies the constraints
(\ref{52}) and (\ref{53}). Since $1/\sigma^2_X+\sum_{i=1}^L
(1-\exp(-2r_{i}))/\sigma^2_{N_i}$ is a strictly concave function
of $r_{\mathcal{I}_L}$, the inequality in (\ref{63}) is strict,
which results in a contradiction with (\ref{equal}).

Now only (\ref{79}) remains to be proved. We shall first show that
$r^*_i(R_{\mathcal{I}_L})=0$ implies $R_i=0$. Without loss of
generality, suppose $r^*_L(R_{\mathcal{I}_L})=0$. Then it is easy
to check that (\ref{52}) still holds if we set $R_L=0$ on its left
hand side. So if $R_L>0$, we can increase
$r^*_L(R_{\mathcal{I}_L})$ by a small amount without violating
(\ref{52}) and (\ref{53}), which is contradictory to the fact that
$r^*_L(R_{\mathcal{I}_L})$ is unique. Hence without loss of
generality, we can assume $r^*_i(R_{\mathcal{I}_L})>0$ for all
$i\in\mathcal{I}_L$. Otherwise by restricting to the set
$\mathcal{G}=\{i\in\mathcal{I}_L:r^*_i(R_{\mathcal{I}_L})>0\}$,
the following argument can still be applied.

Since (\ref{78}) holds, the righthand side of (\ref{52}) becomes
$f(\mathcal{A},r^*_{\mathcal{I}_L}(R_{\mathcal{I}_L}))$. By
(\ref{41}), it can be shown that if in (\ref{52}), the constraints
on $\sum_{i\in\mathcal{S}}R_i$ and $\sum_{i\in\mathcal{T}}R_i$ are
tight, then either $\mathcal{S}\subseteq\mathcal{T}$ or
$\mathcal{T}\subseteq\mathcal{S}$. Let
$\mathcal{\widetilde{A}}=\bigcup_{k\in\mathcal{I}_K}\mathcal{A}_k$,
where $\mathcal{A}_k (k\in\mathcal{I}_K)$ are the sets for which
the constraints on $\sum_{i\in\mathcal{A}_k}R_i$ are tight in
(\ref{52}). If there is no such an $\mathcal{A}_k$, let
$\mathcal{\widetilde{A}}=\emptyset$.

If $\mathcal{\widetilde{A}}=\mathcal{I}_L$, we are done. Otherwise
pick any $i^*\in\mathcal{I}_L\backslash\mathcal{\widetilde{A}}$.
We can increase $r^*_{i^*}(R_{\mathcal{I}_L})$ to
$r^*_{i^*}(R_{\mathcal{I}_L})+\delta$ for some $\delta>0$ without
violating any constraints in (\ref{128}) and (\ref{69}), which is
contradictory to the uniqueness of $r^*_{i^*}(R_{\mathcal{I}_L})$.

\end{document}